\documentclass{IEEEtran}
\usepackage{amsmath,graphicx,cite}
\usepackage[psamsfonts]{amssymb}
\usepackage[OT2,T1]{fontenc}
\usepackage[russian,english]{babel}
\usepackage[colorlinks, bookmarks=false]{hyperref}
\newcommand{\Li}{\operatorname{Li}}
\newcommand{\tr}{\operatorname{tr}}
\newtheorem{lem}{Lemma}
\begin{document}
\title{On the Second-Order Statistics of the Instantaneous
Mutual Information in Rayleigh Fading Channels}
\author{\authorblockN{
Shuangquan Wang, \IEEEmembership{Student Member, IEEE, }Ali Abdi,
\IEEEmembership{Member, IEEE, }}
\thanks{This paper was presented in part at the
\emph{IEEE 6th Workshop on Signal Processing
 Advances in Wireless Communications}, New York City, NY, 2005.}%
\thanks{S.~Wang and A.~Abdi are with the Center for Wireless Communications and Signal
Processing Research (CWCSPR), Department of Electrical and Computer
Engineering, New Jersey Institute of Technology, Newark, NJ 07102,
USA (e-mail:\{sw27, ali.abdi\}@njit.edu).}}
\def\citepunct{][}
\def\citedash{]--[}
\maketitle
\begin{abstract}
In this paper, the second-order statistics of the instantaneous
mutual information are studied, in time-varying Rayleigh fading
channels, assuming general non-isotropic scattering environments.
Specifically, first the autocorrelation function, correlation
coefficient, level crossing rate, and the average outage duration of
the instantaneous mutual information are investigated in
single-input single-output (SISO) systems. Closed-form exact
expressions are derived, as well as accurate approximations in low-
and high-SNR regimes. Then, the results are extended to
multiple-input single-output and single-input multiple-output
systems, as well as multiple-input multiple-output systems with
orthogonal space-time block code transmission. Monte Carlo
simulations are provided to verify the accuracy of the analytical
results. The results shed more light on the dynamic behavior of the
instantaneous mutual information in mobile fading channels.
\end{abstract}

\section{Introduction}\label{sec:introduction}
\PARstart{T}{he} increasing demand for wireless communication over
time-varying channels has motivated further investigation of the
channel dynamics and its statistical behavior. There are numerous
studies on the temporal second-order characteristics of a variety of
terrestrial \cite{IEEE_sw27:JakesBook94,
IEEE_sw27:Abdi00,IEEE_sw27:Youssef96,IEEE_sw27:Abdi02_AoA} and
satellite channels~\cite{IEEE_sw27:Abdi03_satellite_model}, such as
the correlation function, level crossing rate (LCR), and average
fade duration.

For such important quantity as the \emph{instantaneous mutual
information} (IMI), however, only the mean value, which is the
ergodic capacity, has received much attention as well as the outage
probability\cite{IEEE_sw27:Ozarow94,IEEE_sw27:TseBook}. Clearly,
ergodic capacity and outage probability do not show the dynamic
temporal behavior of IMI in time-varying fading channels. For
example, the outage probability of a fading channel gives the
probability of IMI to be smaller than a particular data
rate\cite{IEEE_sw27:Ozarow94}. Nevertheless, it does not show how
long, on average, IMI stays below that rate. To the best of our
knowledge, only some simulation results regarding such second-order
statistics are given in\cite{IEEE_sw27:Giorgetti03}, without
analytical results, and in \cite{IEEE_sw27:NanZhang05}, lower and
upper bounds on the correlation coefficient in the high
signal-to-noise ratio (SNR) regime, as well as some approximations,
are derived without exact results.

In this paper, several second-order statistics of IMI are studied in
time-varying Rayleigh flat fading channels, considering a general
non-isotropic scattering propagation environment. Closed-form
expressions and simple approximations are derived for the
autocorrelation function (ACF), correlation coefficient, LCR and the
average outage duration (AOD) of IMI. A variety of channels are
considered such as multiple-input single-output (MISO), single-input
multiple-output (SIMO), multiple-input multiple-output with
orthogonal space-time block code (OSTBC) transmission, and MIMO in
the low-SNR regime. Monte Carlo simulations are provided to verify
the accuracy of our closed-form expressions and approximate results.

{\it Notation}: ${}^{\dag}$ is reserved for matrix Hermitian,
${}^{\star}$ for complex conjugate, $\tr[\cdot]$ for the trace of a
square matrix, $\jmath$ for $\sqrt{-1}$, $\mathbb{E}[\cdot]$ for
mathematical expectation, $\mathbf{I}_m$ for the $m\times m$
identity matrix, $\ln(\cdot)$ for the natural logarithm,
$\log_2(\cdot)$ for the base-2 logarithm, $\mathcal{O}(\cdot)$ for
the order, $\|\cdot\|_F$ for the Frobenius norm, $f^2(x)$ for
$\left[f(x)\right]^2$, and $\max_{a\leq x\leq b}f(x)$ for the
maximum of $f(x)$ over the range $a\leq x\leq b$.

The rest of this paper is organized as follows. Sec.
\ref{sec:IMI_SISO} introduces the IMI random process in a
single-input single-output (SISO) system, whereas the channel and
angle-of-arrival (AoA) models are described in Appendix
\ref{app:Channel_AoA_Models}. Sec. \ref{sec:ACF_Coeff_SISO} is
devoted to the derivation of ACF and the correlation coefficient of
IMI in SISO channels, as well as their low- and high-SNR
approximations. The LCR and AOD of SISO IMI are derived in Sec.
\ref{sec:LCR_AOD_SISO}. Extension of the SISO results systems with
multiple antennas are discussed in Sec. \ref{sec:OSTBC}. Numerical
results are presented in Sec.
\ref{sec:Numerical_Results_Discussion}, and concluding remarks are
given in Sec. \ref{sec:conclusion}.

\section{The SISO IMI}\label{sec:IMI_SISO}
Similar to \cite{IEEE_sw27:Ozarow94}, we consider a piecewise
constant approximation for the continuous-time SISO fading channel
coefficient $h(t)$, represented by
$\left\{h(lT_{\!s})\right\}_{l=1}^L$, where $T_{\!s}$ is the symbol
duration and $L$ is the number of samples. In the presence of
additive white Gaussian noise, if perfect channel state information
$\left\{h(lT_{\!s})\right\}_{l=1}^L$ is available at the receiver
only, the ergodic channel capacity is given
by\cite{IEEE_sw27:Ozarow94,IEEE_sw27:TseBook}
\begin{equation}\label{eq:Cap_SISO}
C=\mathbb{E}\left[\log_2\left(1+\eta\alpha_l^2\right)\right],
\end{equation} bps/Hz, where $\eta$ is the average SNR at the
receiver side, $\alpha_l$ is the envelop of $h(lT_{\!s})$, i.e.,
$\alpha_l=|h(lT_{\!s})|, \forall l$. We choose
$\mathbb{E}[\alpha_l^2]=1$, i.e., the channel has unit power.

In the above equation, at any given time index $l$,
$\log_2(1+\eta\alpha^2_l)$ is a random variable as it depends on the
fading parameter $\alpha_l$\cite{IEEE_sw27:Ozarow94}. Therefore
\begin{equation}\label{eq:Def_IMI_SISO}
\mathcal{I}_l=\log_2\left(1+\eta\alpha^2_l\right), \quad
l=1,2,\cdots,
\end{equation} is a discrete-time random process with
the ergodic capacity as its mean. We study the second-order
statistics of $\{\mathcal{I}_l\}_{l=1}^\infty$, such as
autocorrelation, correlation coefficient, LCR and AOD, in the
following sections.
\section{ACF and Correlation Coefficient\\of the SISO IMI}
\label{sec:ACF_Coeff_SISO} In this section, first we concentrate on
the ACF of IMI, define in (\ref{eq:Def_IMI_SISO}). The ACF is
defined by
\begin{align}
r_\mathcal{I}(i)&\!=\!\mathbb{E}[\mathcal{I}_l\mathcal{I}_{l-i}],
\label{eq:Def_ACF_SISO}\\
&\!=\!(\log_2e)^2\mathbb{E}\!\left[\ln(1\!+\!\eta x_1)
\!\ln(1\!+\!\eta x_2)\right],\label{eq:ACF_General_SISO}
\end{align} where $x_1=\alpha^2_l$, $x_2=\alpha^2_{l-i}$, and
$x_1$ and $x_2$ have a joint Chi-square probability density function
(PDF) with 2 degrees of
freedom\cite[(3.2)]{IEEE_sw27:Ozarow94}\cite[pp.~163,
(8-103)]{IEEE_sw27:DavenportBook87}
\begin{equation}\label{eq:jpdf_x1_x2_SISO}
p(x_1,x_2)=\lambda_ie^{-\lambda_i(x_1+x_2)}
I_0\!\left(2\lambda_i\varrho_i\sqrt{x_1x_2}\right),
\end{equation} where $\lambda_i=\frac{1}{1-\varrho_i^2}$,
$\varrho_i=|\rho_h(iT_{\!s})|<1$, $i\neq0$,
$\rho_h(\tau)=\mathbb{E}[h(t)h^\star(t-\tau)]$ is the correlation
coefficient of the Rayleigh fading channel, and
$I_k(z)=\frac{1}{\pi}\int_0^\pi e^{z\cos\theta}
\cos(k\theta)\mathrm{d}\theta$ is the $k^\text{th}$ order modified
Bessel function of the first kind. A closed-form expression for
$\rho_h(\tau)$ in non-isotropic scattering environments is given in
the Appendix \ref{app:Channel_AoA_Models}.

Combining (\ref{eq:ACF_General_SISO}) and
(\ref{eq:jpdf_x1_x2_SISO}), with the following Taylor series
\cite[pp. 971, 8.447.1]{IEEE_sw27:RyzhikBook_5th}
\begin{equation}\label{eq:BesselI0_SeriesRep}
I_0(t)=\sum_{k=0}^{\infty}\frac{t^{2k}}{(k!)^22^{2k}},
\end{equation} simplifies (\ref{eq:ACF_General_SISO}) to
the following exact infinite-summation closed-form representation
\begin{equation}\label{eq:ACF_InfSum_SISO}
\begin{split}
\frac{r_\mathcal{I}(i)}{(\log_2e)^2}\!&=\!\lambda_i\!
\sum_{k=0}^{\infty}\!\frac{(\lambda_i\varrho_i)^{2k}}{(k!)^2}
\!\left[\int_0^\infty \!\!\!\!x^k\ln(1\!+\!\eta x)e^{-\lambda_i
x}\mathrm{d}x\right]^2\\
\!&=\!\!\frac{1}{\lambda_i}
\sum_{k=0}^{\infty}\!\left[\frac{\varrho_i^k}{k!}G_{2, 3}^{3,
1}\!\!\left(\!\frac{\lambda_i}{\eta}\!\left|\!\!
\begin{array}{c}
0,1 \\
0,0,k\!+\!1
\end{array}\right.
\!\!\!\!\right)\!\!\right]^{\!2}\!\!\!\!,
\end{split}
\end{equation} where $G$ is
Meijer's $G$ function\cite[pp.~1096,
9.301]{IEEE_sw27:RyzhikBook_5th} and the normalized ACF is given by
\begin{equation}\label{eq:NACF_InfSum_SISO}
\tilde{r}_\mathcal{I}(i)=\frac{r_\mathcal{I}(i)}
{\mathbb{E}[\mathcal{I}_l^2]}=\frac{
{\displaystyle\frac{1}{\lambda_i}\sum_{k=0}^{\infty}}\!
\left[\frac{\varrho_i^k}{k!} G_{2, 3}^{3,
1}\!\!\left(\!\frac{\lambda_i}{\eta}\!\left|\!\!
\begin{array}{c}
0,1 \\
0,0,k\!+\!1
\end{array}\right.
\!\!\!\!\right)\!\right]^{\!2}}{2e^{\frac{1}{\eta}}
G_{2,3}^{3,0}\!\!\left(\frac{1}{\eta}\!\left|\!\!\begin{array}{c}
 1,1 \\
 0,0,0
\end{array}\right.
\!\!\right)}.
\end{equation}

After some algebraic manipulations, we obtain the correlation
coefficient as
\begin{align}
\rho_\mathcal{I}(i)&=\frac{r_\mathcal{I}(i)-\{\mathbb{E}[\mathcal{I}_l]\}^2}
{\mathbb{E}[\mathcal{I}_l^2]-\{\mathbb{E}[\mathcal{I}_l]\}^2},
\label{eq:Def_Coeff_SISO}\\
&\hspace{-1em}=\frac{
{\displaystyle\frac{1}{\lambda_i}\sum_{k=0}^{\infty}}
\!\left[\frac{\varrho_i^k}{k!} G_{2, 3}^{3,
1}\!\!\left(\!\frac{\lambda_i}{\eta}\!\left|\!\!
\begin{array}{c}
0,1 \\
0,0,k\!+\!1
\end{array}\right.
\!\!\!\!\right)\!\right]^{\!2}\!\!-\!
e^{\frac{2}{\eta}}\Gamma^2\!\left(0,\frac{1}{\eta}\right)}{2e^{\frac{1}{\eta}}
G_{2,3}^{3,0}\!\!\left(\frac{1}{\eta}\!\left|\!\!\begin{array}{c}
 1,1 \\
 0,0,0
\end{array}\right.
\!\!\right)-e^{\frac{2}{\eta}}\Gamma^2\!\left(0,\frac{1}{\eta}\right)},
\label{eq:Coeff_InfSum_SISO}
\end{align} where $\Gamma(a,z) =\int_z^\infty
t^{a-1}e^{-t}\mathrm{d}t$\cite[pp.~949,
8.350.2]{IEEE_sw27:RyzhikBook_5th} is the upper incomplete gamma
function. The derivation of (\ref{eq:ACF_InfSum_SISO}),
(\ref{eq:NACF_InfSum_SISO}) and (\ref{eq:Coeff_InfSum_SISO}) are
given in Appendix \ref{app:SISO_IMI:sec:ACF_Coeff}.

In general, it seems difficult to further simplify
(\ref{eq:NACF_InfSum_SISO}) and (\ref{eq:Coeff_InfSum_SISO}).
However, we note that the integral
$\Xi(k,\eta,\lambda_i)=\int_0^\infty x^ke^{-\lambda_ix}\ln(1+\eta
x)\mathrm{d}x$, $k\geq0$, can be approximated by
\begin{equation}\label{eq:Xi_LowHighSNR_SISO}
\Xi(k,\eta,\lambda_i)\approx
\begin{cases}
\int_0^\infty \eta x^{k+1}e^{-\lambda_ix}\mathrm{d}x,& \eta\rightarrow0,\\
\int_0^\infty x^ke^{-\lambda_ix}\ln(\eta x)\mathrm{d}x,&
\eta\rightarrow\infty,
\end{cases}
\end{equation} using $\ln(1+\eta x)\approx\eta x$, $\eta\rightarrow0$, and
$\ln(1+\eta x)\approx\ln\left(\eta x\right)$,
$\eta\rightarrow\infty$, respectively\footnote{The utility and
accuracy of (\ref{eq:Xi_LowHighSNR_SISO}) is confirmed by Monte
Carlo simulations in Sec. \ref{sec:Numerical_Results_Discussion}.}.

In the following two subsections, we will use
(\ref{eq:ACF_InfSum_SISO}) and (\ref{eq:Xi_LowHighSNR_SISO}) to
derive asymptotic closed-form expressions for $r_\mathcal{I}(i)$ in
the low- and high-SNR regimes.

\subsection{Low-SNR Regime}
\label{ssec:ACF_Coeff_LowSNR_SISO} If $\eta\rightarrow0$, based on
(\ref{eq:ACF_InfSum_SISO}) and (\ref{eq:Xi_LowHighSNR_SISO}), we
have
\begin{equation}\label{eq:ACF_XiL_LowSNR_SISO}
\frac{r_\mathcal{I}(i)}{(\log_2e)^2}\!\approx\!\lambda_i\!
\sum_{k=0}^{\infty}\frac{\Xi_L^2(k,\eta,\lambda_i)}
{(\lambda_i\varrho_i)^{-2k}(k!)^2},
\end{equation} where $\Xi_L(k,\eta,\lambda_i)$ is
defined as $\eta\!\int_0^\infty \!\!x^{k+1}e^{-\lambda_i
x}\mathrm{d}x$ and given by\cite[pp.~364,
3.381.4]{IEEE_sw27:RyzhikBook_5th}
\begin{equation}\label{eq:Xi_L}
\Xi_L(k,\eta,\lambda_i)=\eta\frac{(k+1)!}{\lambda_i^{k+2}}.
\end{equation} By replacing $\Xi_L(k,\eta,\lambda_i)$ in
(\ref{eq:ACF_XiL_LowSNR_SISO}) with (\ref{eq:Xi_L}), we obtain
\begin{equation}\label{eq:ACF_InfSum_LowSNR_SISO}
\begin{split}\frac{r_\mathcal{I}(i)}{(\log_2e)^2}\!&\approx\!\lambda_i\!
\sum_{k=0}^{\infty}\!\frac{(\lambda_i\varrho_i)^{2k}}{(k!)^2}
\!\left[\!\eta\frac{(k+1)!}{\lambda_i^{k+2}}\right]^2,\\
\!&=\!\!\frac{\eta^2}{\lambda_i^3\varrho_i^2}\!
\sum_{k=1}^{\infty}\!k^2\varrho_i^{2k}.
\end{split}
\end{equation} Since $\varrho_i<1$, we have
\begin{equation}\label{eq:InfSum_k2rho2}
\sum_{k=1}^{\infty}\!k^2\varrho_i^{2k}=\left.t\frac{\partial}
{\partial t}\left[t\left(\frac{\partial} {\partial
t}\sum_{k=1}^\infty t^k\right)\right]\right|_{t=\varrho_i^2}
\!\!\!\!=\lambda_i^3\varrho_i^2(1+\varrho_i^2),
\end{equation} which simplifies
(\ref{eq:ACF_InfSum_LowSNR_SISO}) to
\begin{equation}\label{eq:ACF_LowSNR_SISO}
r_\mathcal{I}(i)\approx(\log_2e)^2\eta^2(1+\varrho_i^2).
\end{equation}

After normalization by $\mathbb{E}[\mathcal{I}_l^2]
=r_\mathcal{I}(0)\approx2(\log_2e)^2\eta^2$, we obtain
\begin{equation}\label{eq:NACF_LowSNR_SISO}
\tilde{r}_\mathcal{I}(i)\approx\frac{1+\varrho_i^2}{2}.
\end{equation}

Moreover, the correlation coefficient is given by
\begin{equation}\label{eq:Coeff_LowSNR_SISO}
\rho_\mathcal{I}(i)\!\approx\!\frac{r_\mathcal{I}(i)\!-
\!\{\!\left(\log_2\!e\!\right)\!\mathbb{E}[\eta
x]\!\}^2}{\mathbb{E}[\mathcal{I}_l^2]\!-
\!\{\!\left(\log_2\!e\!\right)\!\mathbb{E}[\eta
x]\!\}^2}\!=\!\frac{\eta^2(1\!+\!\varrho_i^2)\!-\!\eta^2}{2\eta^2\!-\!\eta^2}
\!=\!\varrho_i^2,
\end{equation} where we use
$\mathbb{E}[x]=\mathbb{E}[\alpha_l^2]=1$, due to the unit-power
Rayleigh fading assumption.

With isotropic scattering, we have $\rho_h(\tau)=J_0(2\pi f_m\tau)$,
the Clarke's correlation\cite{IEEE_sw27:JakesBook94}, where $f_m$ is
the maximum Doppler frequency, and $J_0(x)=\frac{1}{\pi}\int_0^\pi
e^{\jmath x\cos\theta}\mathrm{d}\theta$ is the zero-order Bessel
function of the first kind. This simplifies
(\ref{eq:Coeff_LowSNR_SISO}) to $J_0^2(2\pi f_miT_{\!s})$ with
$\tau=iT_{\!s}$.
\subsection{High-SNR Regime}
\label{ssec:ACF_Coeff_HighSNR_SISO} If $\eta\rightarrow\infty$,
based on (\ref{eq:ACF_InfSum_SISO}) and
(\ref{eq:Xi_LowHighSNR_SISO}), we get
\begin{equation}\label{eq:ACF_XiH_HighSNR_SISO}
\frac{r_\mathcal{I}(i)}{(\log_2e)^2}\!\approx\!\lambda_i\!
\sum_{k=0}^{\infty}\frac{\Xi_H^2(k,\eta,\lambda_i)}
{(\lambda_i\varrho_i)^{-2k}(k!)^2},
\end{equation} where $\Xi_H(k,\eta,\lambda_i)$ is defined as $\int_0^\infty
\!\!x^k(\ln\eta+\ln x)e^{-\lambda_i x}\mathrm{d}x$. Using
4.352.2~\cite[pp.~604]{IEEE_sw27:RyzhikBook_5th} and
(\ref{eq:Xi_L}), we obtain
\begin{equation}\label{eq:Xi_H}
\Xi_H(k,\eta,\lambda_i)=\frac{k!}{\lambda_i^{k+1}}\left(\ln\frac{\eta}{\lambda_i
\gamma}+{H}_{\!k}\right),
\end{equation} where $\gamma=1.781072\cdots$ is the Euler-Mascheroni constant
\cite[pp.~\textrm{xxx}]{IEEE_sw27:RyzhikBook_5th}, ${H}_{\!k}$ is
the $k^\text{th}$ harmonic number\cite[pp. 29,
(2.13)]{IEEE_sw27:Knuth_2nd}, defined by
${H}_{\!k}=\sum_{j=1}^k\frac{1}{j}$ for $k\geq1$, and ${H}_0=0$.
After lengthy algebraic calculations, finally the normalized ACF in
the high-SNR regime is shown to be
\begin{equation}\label{eq:NACF_HighSNR_SISO}
\tilde{r}_\mathcal{I}(i)\approx\frac{\Li_2\!\left(\varrho_i^2\right)
+\ln^2\!\frac{\eta}{\gamma}}
{\frac{\pi^2}{6}+\ln^2\!\frac{\eta}{\gamma}},
\end{equation} where $\Li_2(x)$ is the dilogarithm function,
defined as $\Li_2(x)=\sum_{k=1}^\infty\frac{x^k}{k^2}, |x|\leq1$.
The correlation coefficient is given by
\begin{equation}\label{eq:Coeff_HighSNR_SISO}
\rho_\mathcal{I}(i)\approx\frac{6\Li_2\!\left(\varrho_i^2\right)}{\pi^2}.
\end{equation}

Derivation of (\ref{eq:NACF_HighSNR_SISO}) and
(\ref{eq:Coeff_HighSNR_SISO}) are given in Appendix
\ref{app:SISO_IMI:sec:ACF_Coeff_High_SNR}. With isotropic
scattering, (\ref{eq:Coeff_HighSNR_SISO}) reduces to
\begin{equation}\label{eq:Coeff_HighSNR_ISO_SISO}
\rho_\mathcal{I}(i)\approx\frac{6\Li_2\!\!\left[J_0^2\!\left(2\pi
f_miT_{\!s}\right)\right]}{\pi^2}.
\end{equation}
\section{LCR and AOD of SISO IMI}
\label{sec:LCR_AOD_SISO}
\subsection{The LCR of SISO IMI}\label{ssec:LCR_SISO} Similar to the
calculation of zero crossing rate in discrete time\cite[Ch.
4]{IEEE_sw27:KedemBook94}, we define the binary sequence
$\left\{X_l\right\}_{l=1}^L$, based on the IMI sequence
$\left\{\mathcal{I}_1\right\}_{l=1}^L$, as
\begin{equation}\label{eq:Hardening_Function}
X_l=\begin{cases} 1,&\text{if } \mathcal{I}_l\geq
\mathcal{I}_\text{th},\\
0,&\text{if } \mathcal{I}_l<\mathcal{I}_\text{th},
\end{cases}
\end{equation} where $\mathcal{I}_\text{th}$ is a fixed threshold. The
number of crossings of $\left\{\mathcal{I}_1\right\}_{l=1}^L$ with
$\mathcal{I}_\text{th}$,  within the time interval $T_{\!s}\leq
t\leq LT_{\!s}$, denoted by $D_{\mathcal{I}_\text{th}}$, is defined
in terms of $\{X_l\}_{l=1}^L$\cite[(4.1)]{IEEE_sw27:KedemBook94}
\begin{equation}\label{eq:Def_Num_UpDownCrossings}
D_{\mathcal{I}_\text{th}}=\sum_{l=2}^L\left(X_l-X_{l-1}\right)^2,
\end{equation} which includes both up- and down-crossings.

After some simple manipulations, the expected crossing rate at the
level $\mathcal{I}_\text{th}$ can be written as
\begin{equation}\label{eq:Def_LCR_UpDown_DT}
\frac{\mathbb{E}[D_{\mathcal{I}_\text{th}}]}{(L-1)T_{\!s}}
=\frac{2P_r\{X_l=1\}-2P_r\{X_l=1,X_{l-1}=1\}}{T_{\!s}}.
\end{equation} Therefore the expected down crossing rate at the level
$\mathcal{I}_\text{th}$, denoted by
$N_\mathcal{I}(\mathcal{I}_\text{th})$, is given by
\begin{equation}\label{eq:Def_LCR_Up_DT}
N_\mathcal{I}(\mathcal{I}_\text{th})=\frac{P_r\{X_l=1\}-P_r\{X_l=1,X_{l-1}=1\}}{T_{\!s}}.
\end{equation} To simplify the notation, we use $\phi$ for
$P_r\{X_l=1\}$ and $\varphi(\varrho_1)$ to denote
$P_r\{X_l=1,X_{l-1}=1\}$, where $\varrho_1=|\rho_h(T_{\!s})|$,
defined before. They can be calculated as follows
\begin{equation}\label{eq:Prob_Xl}
\phi=P_r\left\{\alpha^2_l\geq\frac{2^{\mathcal{I}_\text{th}}-1}{\eta}\right\}
=\int_\frac{2^{\mathcal{I}_\text{th}}-1}{\eta}^\infty
e^{-x}\mathrm{d}x=e^{-\frac{2^{\mathcal{I}_\text{th}}-1}{\eta}},
\end{equation}
\begin{equation}\label{eq:Prob_Xl_Xl-1}
\begin{split}
\varphi(\varrho_1)&=P_r\left\{\alpha^2_l\geq\frac{2^{\mathcal{I}_\text{th}}-1}
{\eta},\alpha^2_{l-1}\geq\frac{2^{\mathcal{I}_\text{th}}-1}{\eta}\right\},\\
&=\lambda_1\sum_{k=0}^\infty\frac{(\lambda_1\varrho_1)^{2k}}
{(k!)^2}\left[\int_\frac{2^{\mathcal{I}_\text{th}}-1}{\eta}^\infty
x^ke^{-\lambda_1x}\mathrm{d}x\right]^2,\\
&=\left(1-\varrho_1^2\right)\sum_{k=0}^\infty\left[\frac{\varrho_1^k}
{k!}\Gamma\!\left(k+1,\frac{2^{\mathcal{I}_\text{th}}-1}
{\eta\left(1-\varrho_1^2\right)}\right)\right]^2.
\end{split}
\end{equation}

By plugging (\ref{eq:Prob_Xl}) and (\ref{eq:Prob_Xl_Xl-1}) into
(\ref{eq:Def_LCR_Up_DT}), we obtain the expected crossing rate at
the level $\mathcal{I}_\text{th}$ as
\begin{equation}\label{eq:LCR_Up_DT_SISO}
N_\mathcal{I}(\mathcal{I}_\text{th})\!\!=\!\!\frac{e^{-\frac{2^{\mathcal{I}_\text{th}}-1}
{\eta}}}{T_{\!s}}-
\frac{1-\varrho_1^2}{T_{\!s}}\!\sum_{k=0}^\infty\!\!\left[\!\frac{\varrho_1^k}
{k!}\Gamma\!\!\left(\!k\!+\!1,\frac{2^{\mathcal{I}_\text{th}}\!-\!1}
{\eta\!\left(1-\varrho_1^2\right)}\!\right)\!\!\right]^2\!\!\!\!.
\end{equation}

\subsection{The AOD of SISO IMI}\label{ssec:AOD_SISO}
Based on the relationship between $\mathcal{I}_l$ and $\alpha_l^2$
in (\ref{eq:Def_IMI_SISO}), the cumulative distribution function
(CDF) of $\mathcal{I}_l$ is obtained as
\begin{equation}\label{eq:CDF_IMI_SISO}
F_\mathcal{I}(\mathcal{\mathcal{I}_\text{th}})=P_r\left\{\alpha^2_l\leq
\frac{2^{\mathcal{I}_\text{th}}-1}{\eta}\right\}=
1-e^{-\frac{2^{\mathcal{I}_\text{th}}-1}{\eta}}.
\end{equation}

The AOD of IMI, normalized by $T_{\!s}$, is therefore given
by\footnote{The definition of IMI AOD is similar to the average
envelope fade duration\cite{IEEE_sw27:JakesBook94}.}
\begin{equation}\label{eq:AOD_SISO}
\begin{split}
\frac{\overline{t}_\mathcal{I}(\mathcal{I}_\text{th})}{T_{\!s}}&=\frac{F_\mathcal{I}(\mathcal{I}_\text{th})}
{T_{\!s}N_\mathcal{I}(\mathcal{I}_\text{th})},\\
&\hspace{-2em}=\frac{1-e^{-\frac{2^{\mathcal{I}_\text{th}}-1}{\eta}}}
{e^{-\frac{2^{\mathcal{I}_\text{th}}-1}{\eta}}-
\left(1-\varrho_1^2\right)\!\sum_{k=0}^\infty\!\!\left[\!\frac{\varrho_1^k}
{k!}\Gamma\!\!\left(\!k\!+\!1,\frac{2^{\mathcal{I}_\text{th}}\!-\!1}
{\eta\left(1-\varrho_1^2\right)}\!\right)\!\!\right]^2}.
\end{split}
\end{equation}
\section{Extension to Multiple Antenna Systems}
\label{sec:OSTBC} For $N$-receiver (Rx) SIMO, $M$-transmitter (Tx)
MISO, and $M$-Tx $N$-Rx OSTBC-based MIMO systems, we assume
perfectly estimated, independent and identically distributed
subchannels, with the same temporal correlation coefficient
$\rho_h(\tau)$. Therefore the IMIs can be written
as\cite{IEEE_sw27:Telatar99_MIMO_Cap,IEEE_sw27:TseBook}\cite[pp.
115, (7.4.43)]{IEEE_sw27:LarssonBook}
\begin{subequations}
\begin{align}
\mathcal{I}^{SIMO}_l&=\log_2\left[1+\eta\sum_{r=1}^N|h_r(lT_{\!s})|^2\right],
\label{eq:Def_IMI_SIMO}\\
\mathcal{I}^{MISO}_l&=\log_2\left[1+\frac{\eta}{M}
\sum_{t=1}^M|h_t(lT_{\!s})|^2\right],\label{eq:Def_IMI_MISO}\\
\mathcal{I}^{OSTBC}_l&=\log_2\left[1+\frac{\eta}{M}
\sum_{r=1}^N\sum_{t=1}^M|h_{r,t}(lT_{\!s})|^2\right]\label{eq:Def_IMI_OSTBC},
\end{align}
\end{subequations}
\begin{figure*}[!t]
\normalsize
\begin{equation}\label{eq:NACF_OSTBC}
\tilde{r}_\mathcal{I}(i)=\frac{\frac{1}{
2e^\frac{M}{\eta}\left(\frac{M\lambda_i}{\eta}\right)^{M\!N}}
{\displaystyle\sum_{k=0}^\infty}\frac{\varrho_i^{2k} \left[G_{2,
3}^{3, 1}\!\left(\!\frac{M\lambda_i}{\eta}\!\left|\!\!
\begin{array}{c}
0,1 \\
0,0,k\!+\!M\!N
\end{array}\right.
\!\!\!\!\right)\right]^2}
{k!(M\!N+k-1)!}}{{\displaystyle\sum_{j=0}^{M\!N-1}}{M\!N-1 \choose
j}(-1)^{M\!N-1-j}G_{3,4}^{4,0}\!\left(\!\frac{M}{\eta}\!\left|
\begin{array}{c}
 -\!j,-\!j,-\!j \\
 0,-\!j\!-\!1,-\!j\!-\!1,-\!j\!-\!1
\end{array}
\right.\!\!\!\!\right)}.
\end{equation}
\hrulefill
\begin{equation}\label{eq:Coeff_OSTBC}
\rho_\mathcal{I}(i)=\frac{\frac{(M\!N-1)!}{\lambda_i^{M\!N}}
{\displaystyle\sum_{k=0}^\infty}\frac{\varrho_i^{2k} \left[G_{2,
3}^{3, 1}\!\left(\!\frac{M\lambda_i}{\eta}\!\left|\!\!
\begin{array}{c}
0,1 \\
0,0,k\!+\!M\!N
\end{array}\right.
\!\!\!\!\right)\right]^2}
{k!(M\!N+k-1)!}-\left[G_{2,3}^{3,1}\!\left(\!\frac{M}{\eta}\!\left|
\begin{array}{c}
 0,1 \\
 0,0,M\!N
\end{array}
\right.\!\!\right)\right]^2}{2(M\!N\!-\!1)!e^\frac{M}{\eta}\left(\frac{M}{\eta}\right)^{M\!N}
 {\displaystyle\sum_{j=0}^{M\!N-1}}{M\!N-1 \choose
j}(-1)^{M\!N-1-j}G_{3,4}^{4,0}\!\left(\!\frac{M}{\eta}\!\left|
\begin{array}{c}
 -\!j,-\!j,-\!j \\
 0,-\!j\!-\!1,-\!j\!-\!1,-\!j\!-\!1
\end{array}
\right.\!\!\!\!\right)
-\left[G_{2,3}^{3,1}\!\left(\!\frac{M}{\eta}\!\left|
\begin{array}{c}
 0,1 \\
 0,0,M\!N
\end{array}
\right.\!\!\right)\right]^2}.
\end{equation}
\hrulefill
\vspace*{4pt}
\end{figure*}
where $\eta$ is the expected SNR at each Rx antenna. Note that the
SIMO and MISO systems in (\ref{eq:Def_IMI_SIMO}) and
(\ref{eq:Def_IMI_MISO}) are known as maximal ratio combiner (MRC)
and maximal ratio transmitter (MRT),
respectively\cite{IEEE_sw27:PaulrajBook_Space_Time}.

It is easy to see (\ref{eq:Def_IMI_OSTBC}) includes
(\ref{eq:Def_IMI_SIMO}) and (\ref{eq:Def_IMI_MISO}) as special
cases, by setting $M=1$ and $N=1$, respectively. Therefore, we only
need to consider the OSTBC-based MIMO system.

At any time index $l$,
$\sum_{r=1}^N\sum_{t=1}^M|h_{r,t}(lT_{\!s})|^2$ is a $\chi^2$ random
variable with the $2M\!N$ degrees of freedom. We define
$y_1=\sum_{r=1}^N\sum_{t=1}^M|h_{r,t}(lT_{\!s})|^2$ and
$y_2=\sum_{r=1}^N\sum_{t=1}^M|h_{r,t}[(l-i)T_{\!s}]|^2$. The PDF of
$y_1$ and the joint PDF of $y_1$ and $y_2$ are, respectively, given
by\cite[(2.32), (3.14)]{IEEE_sw27:SimonBook02}
\begin{align}
p(y_1)&=\frac{y_1^{M\!N-1}e^{-y_1}}{(M\!N-1)!},\label{eq:pdf_y1_OSTBC}\\
p(y_1,y_2)&=\frac{\lambda_i(y_1y_2)^\frac{M\!N-1}{2}
I_{M\!N-1}\!\left(2\lambda_i\varrho_i\sqrt{y_1y_2}\right)}
{(M\!N-1)!\varrho_i^{M\!N-1}e^{\lambda_i(y_1+y_2)}}.
\label{eq:jpdf_y1_y2_OSTBC}
\end{align}

According to the series representation of the modified Bessel
function of the first kind\cite[pp. 971,
8.445]{IEEE_sw27:RyzhikBook_5th}, we can rewrite
(\ref{eq:jpdf_y1_y2_OSTBC}) as
\begin{equation}\label{eq:jpdf_y1_y2_SeriesRep_OSTBC}
p(y_1,y_2)=\frac{\lambda_i^{M\!N} (y_1y_2)^{M\!N-1}}
{(M\!N-1)!e^{\lambda_i(y_1+y_2)}}
\sum_{k=0}^{\infty}\frac{(\lambda_i\varrho_i)^{2k}
(y_1y_2)^k}{k!(M\!N+k-1)!}.
\end{equation}
\subsection{ACF and Correlation Coefficient}
\label{ssec:ACF_Coeff_OSTBC} By substituting
(\ref{eq:Def_IMI_OSTBC}), (\ref{eq:pdf_y1_OSTBC}) and
(\ref{eq:jpdf_y1_y2_OSTBC}) into (\ref{eq:Def_ACF_SISO}),
(\ref{eq:NACF_InfSum_SISO}) and (\ref{eq:Def_Coeff_SISO}), after
some manipulations, we obtained the normalized ACF and the
correlation coefficient as shown in (\ref{eq:NACF_OSTBC}) and
(\ref{eq:Coeff_OSTBC}), respectively, whose derivations are given in
Appendix \ref{app:Deriv_ACF_Coeff_OSTBC}. Note that ${m \choose
n}=\frac{m!}{(m-n)!n!}$, $m\geq n$, in (\ref{eq:NACF_OSTBC}) and
(\ref{eq:Coeff_OSTBC}).

\subsubsection{Low-SNR Regime}
\label{sssec:ACF_Coeff_LowSNR_OSTBC} In Appendix
\ref{app:Deriv_ACF_Coeff_OSTBC_MIMO_Low_SNR}, the normalized ACF and
the correlation coefficient are, respectively, expressed as
\begin{equation}\label{eq:NACF_LowSNR_OSTBC}
\tilde{r}_\mathcal{I}(i)\approx\frac{M\!N+\varrho_i^2}{M\!N+1},
\end{equation} and
\begin{equation}\label{eq:Coeff_LowSNR_OSTBC}
\rho_\mathcal{I}(i)\approx\varrho_i^2.
\end{equation}

As expected, (\ref{eq:NACF_LowSNR_OSTBC}) reduces to
(\ref{eq:NACF_LowSNR_SISO}), when $M=N=1$. Interestingly, the
correlation coefficient in the OSTBC-MIMO system is the same as that
in the SISO system.
\subsubsection{High-SNR Regime}
\label{sssec:ACF_Coeff_HighSNR_OSTBC} The normalized ACF is shown to
be
\begin{multline}\label{eq:NACF_HighSNR_OSTBC}
  \tilde{r}_\mathcal{I}(i)\approx\frac{\frac{(1-\varrho_i^2)^{M\!N}
  R_2(\varrho_i^2)}{(M\!N-1)!}+
\left({H}_{\!M\!N-1}+\ln\frac{\eta}{M\gamma}\right)^2}
{\left({H}_{\!M\!N-1}+\ln\frac{\eta}{M\gamma}\right)^2+\zeta(2,M\!N)}\\
 -\frac{\left[{H}_{\!M\!N-1}-\ln(1-\varrho_i^2)\right]^2}
{\left({H}_{\!M\!N-1}+\ln\frac{\eta}{M\gamma}\right)^2+\zeta(2,M\!N)},
\end{multline} where $\zeta(z,q)=\sum_{k=0}^\infty\frac{1}{(q+k)^z}$
\cite[pp.~1101, 9.521.1]{IEEE_sw27:RyzhikBook_5th} and
$R_2(\varrho_i^2)$ is given in (\ref{eq:R2}). Therefore, the
correlation coefficient is
\begin{equation}\label{eq:Coeff_HighSNR_OSTBC}
  \rho_\mathcal{I}(i)\approx\frac{\frac{(1-\varrho_i^2)^{M\!N}
  R_2(\varrho_i^2)}{(M\!N-1)!}-\left[{H}_{\!M\!N-1}-\ln(1-\varrho_i^2)\right]^2}
  {\zeta(2,M\!N)}.
\end{equation} Derivation of (\ref{eq:NACF_HighSNR_OSTBC})
and (\ref{eq:Coeff_HighSNR_OSTBC}) can be found in Appendix
\ref{app:Deriv_ACF_Coeff_OSTBC_MIMO_High_SNR}. We have checked that
for $M=N=1$, (\ref{eq:NACF_HighSNR_OSTBC}) and
(\ref{eq:Coeff_HighSNR_OSTBC}) simplify to
(\ref{eq:NACF_HighSNR_SISO}) and (\ref{eq:Coeff_HighSNR_SISO}),
respectively, using $R_2(\varrho_i^2)=S_2(\varrho_i^2)$ in
(\ref{eq:S2}).

To obtain some insight, we consider a $2\times2$ OSTBC-MIMO system
in the high-SNR regime. For $M=N=2$, (\ref{eq:R0}), (\ref{eq:R1})
and (\ref{eq:R2}) reduce to
\begin{subequations}\label{eq:R012_M2N2}
\begin{align}
R_0(\varrho_i^2)&=\frac{6}{(1-\varrho_i^2)^4},\\
R_1(\varrho_i^2)&=\frac{11-6\ln(1-\varrho_i^2)}
{(1-\varrho_i^2)^4},\\
R_2(\varrho_i^2)
&\!=\!\frac{6\ln^2(1\!-\!\varrho_i^2)\!+\!6\Li_2\!\left(\varrho_i^2\right)}
{(1-\varrho_i^2)^4}\!+\!\frac{2\left(3\varrho_i^4
\!+\!4\varrho_i^2\!-\!1\right)}{\varrho_i^4(1-\varrho_i^2)^4}\nonumber\\
&\hspace{1em}-\frac{\left(11\varrho_i^6
+18\varrho_i^4-9\varrho_i^2+2\right) \ln
(1-\varrho_i^2)}{\varrho_i^6(1-\varrho_i^2)^4}\label{eq:R2_M2N2}.
\end{align}
\end{subequations} To derive (\ref{eq:R2_M2N2}) from (\ref{eq:R2}), we need the
integral representation of $\Li_2(t)$ as $\Li_2(t)=-\int_0^t
\frac{\ln(1-z)}{z}\mathrm{d}z$, which can be derived from 0.241.4 in
\cite[pp.~12]{IEEE_sw27:RyzhikBook_5th}, by using
$\int_0^1\frac{\ln(1-z)}{z}\mathrm{d}z=-\frac{\pi^2}{6}$\cite[pp.~586,
4.291.2]{IEEE_sw27:RyzhikBook_5th}.
\begin{figure*}[!t]
\normalsize
\begin{equation}\label{eq:AOD_OSTBC}
\frac{\overline{t}_\mathcal{I}(\mathcal{I}_{th})}{T_{\!s}}
=\frac{\gamma\!\left(M\!N, \frac{2^{\mathcal{I}_\text{th}}-1}
{\eta/M}\right)} {\Gamma\!\left(M\!N,
\frac{2^{\mathcal{I}_\text{th}}-1}{\eta/M}\right)-
\left(1-\varrho_1^2\right)^{M\!N}{\displaystyle\sum_{k=0}^\infty}
\frac{\varrho_1^{2k}}{k!(M\!N+k-1)!}\Gamma^2\!\left(k+M\!N,
\frac{M\left(2^{\mathcal{I}_\text{th}}-1\right)}
{\eta\left(1-\varrho_1^2\right)}\right)}.
\end{equation}
\hrulefill
\vspace*{4pt}
\end{figure*}

Using (\ref{eq:R2_M2N2}), (\ref{eq:NACF_HighSNR_OSTBC}) and
(\ref{eq:Coeff_HighSNR_OSTBC}) reduce to, respectively,
\begin{equation}\label{eq:NACF_HighSNR_M2N2}
\begin{split}
\tilde{r}_\mathcal{I}(i)&\approx\frac{6+22\ln\frac{\eta}{2\gamma}
+6\ln^2\frac{\eta}{2\gamma}+6\Li_2\!\left(\varrho_i^2\right)}
{12+\pi^2+22\ln\frac{\eta}{2\gamma}
+6\ln^2\frac{\eta}{2\gamma}}\\
&\hspace{-2em}+\frac{\left(11\varrho_i^6
-18\varrho_i^4+9\varrho_i^2-2\right)\ln(1-\varrho_i^2)
+8\varrho_i^4-2\varrho_i^2}
{\left(12+\pi^2+22\ln\frac{\eta}{2\gamma}
+6\ln^2\frac{\eta}{2\gamma}\right)\varrho_i^6},
\end{split}
\end{equation}
and
\begin{equation}\label{eq:Coeff_HighSNR_M2N2}
\begin{split}
\rho_\mathcal{I}(i)&\approx\frac{36\Li_2\!\left(\varrho_i^2\right)}
{(6\pi^2-49)} -\frac{85\varrho_i^4-48\varrho_i^2+12}
{(6\pi^2-49)\varrho_i^4}\\
&\hspace{1em}+\frac{6\left(11\varrho_i^6
-18\varrho_i^4+9\varrho_i^2-2\right)\ln(1-\varrho_i^2)}
{(6\pi^2-49)\varrho_i^6}.
\end{split}
\end{equation}
\subsection{LCR and AOD}
\label{ssec:LCR_AOD_OSTBC}
\subsubsection{The LCR}\label{sssec:LCR_OSTBC} We set $\beta_l^2=\sum_{r=1}^N
\sum_{t=1}^M|h_{r,t}(lT_{\!s})|^2$, which has a $\chi^2$
distribution with $2M\!N$ degrees of freedom, with the PDF given by
(\ref{eq:pdf_y1_OSTBC}). In addition, $\beta_l^2$ and
$\beta_{l-1}^2$ has the joint PDF given in
(\ref{eq:jpdf_y1_y2_OSTBC}), with $i=1$.  Similar to the SISO case,
$\phi$ and $\varphi(\varrho_1)$ are calculated as
\begin{equation}\label{eq:Prob_Xl_OSTBC}
\begin{split}
\phi&=P_r\left\{\beta^2_l\geq\frac{2^{\mathcal{I}_\text{th}}-1}
{\eta/M}\right\},\\
&=\int_\frac{2^{\mathcal{I}_\text{th}}-1}{\eta/M}^\infty
\frac{y^{M\!N-1}e^{-y}}{(M\!N-1)!}\mathrm{d}y
=\frac{\Gamma\!\left(M\!N,\frac{2^{\mathcal{I}_\text{th}}-1}
{\eta/M}\right)}{(M\!N-1)!},
\end{split}
\end{equation} and
\begin{align}\label{eq:Prob_Xl_Xl-1_OSTBC}
\varphi(\varrho_1)&=P_r\left\{\beta^2_l\geq\frac{2^{\mathcal{I}_\text{th}}-1}
{\eta/M},\beta^2_{l-1}\geq\frac{2^{\mathcal{I}_\text{th}}-1}
{\eta/M}\right\},\nonumber\\
&=\frac{\lambda_1^{M\!N}}{(M\!N-1)!}
\sum_{k=0}^\infty\Bigg\{\frac{(\lambda_1\varrho_1)^{2k}}
{k!(M\!N+k-1)!}\nonumber\\
&\
\times\left[\int_\frac{2^{\mathcal{I}_\text{th}}-1}{\eta/M}^\infty
y^{k+M\!N-1}e^{-\lambda_1y}\mathrm{d}y\right]^2\Bigg\},\nonumber\\
&\hspace{-2em}=\frac{\left(1-\varrho_1^2\right)^{M\!N}}{(M\!N-1)!}\sum_{k=0}^\infty
\frac{\left[\varrho_1^k\Gamma\!\left(k+M\!N,
\frac{M\left(2^{\mathcal{I}_\text{th}}-1\right)}
{\eta\left(1-\varrho_1^2\right)}\right)\right]^2}{k!(M\!N+k-1)!},
\end{align} where the last expression in (\ref{eq:Prob_Xl_Xl-1_OSTBC})
comes from (\ref{eq:jpdf_y1_y2_SeriesRep_OSTBC}).

By plugging (\ref{eq:Prob_Xl_OSTBC}) and
(\ref{eq:Prob_Xl_Xl-1_OSTBC}) into (\ref{eq:Def_LCR_Up_DT}), we
obtain the expected crossing rate of IMI, at the level
$\mathcal{I}_\text{th}$, for the OSTBC-MIMO system as
\begin{multline}\label{eq:LCR_DT_Up}
N_\mathcal{I}(\mathcal{I}_\text{th})=\frac{\Gamma\!\left(M\!N,
\frac{2^{\mathcal{I}_\text{th}}-1}{\eta/M}\right)}{(M\!N-1)!T_{\!s}}-
\frac{\left(1-\varrho_1^2\right)^{M\!N}}{(M\!N-1)!T_{\!s}}\\
\times\sum_{k=0}^\infty \frac{\left[\varrho_1^k\Gamma\!\left(k+M\!N,
\frac{M\left(2^{\mathcal{I}_\text{th}}-1\right)}
{\eta\left(1-\varrho_1^2\right)}\right)\right]^2}{k!(M\!N+k-1)!},
\end{multline} which includes (\ref{eq:LCR_Up_DT_SISO})
as a special case, when $M=N=1$.
\subsubsection{The AOD}\label{sssec:AOD_OSTBC}
Based on the connection between $\mathcal{I}_l^\text{OSTBC}$ and
$\beta_l^2=\sum_{r=1}^N \sum_{t=1}^M|h_{r,t}(lT_{\!s})|^2$ in
(\ref{eq:Def_IMI_OSTBC}) and using (\ref{eq:Prob_Xl_OSTBC}), the CDF
of $\mathcal{I}_l^\text{OSTBC}$ can be written as
$F_\mathcal{I}(\mathcal{I}_\text{th})=\gamma\!\left(M\!N,
\frac{2^{\mathcal{I}_\text{th}}-1} {\eta/M}\right)/(M\!N-1)!$, where
$\gamma(a,z)=\Gamma(a)-\Gamma(a,z)
=\int_0^zt^{a-1}e^{-t}\mathrm{d}t$\cite[pp.~949,
8.350.1]{IEEE_sw27:RyzhikBook_5th} is the lower incomplete gamma
function. The AOD, normalized by $T_{\!s}$, is given in
(\ref{eq:AOD_OSTBC}), where
$\Gamma^2(a,z)=\left[\Gamma(a,z)\right]^2$. By replacing $M$ and $N$
with 1 in (\ref{eq:AOD_OSTBC}), we obtain (\ref{eq:AOD_SISO}).

\section{Numerical Results and Discussion}
\label{sec:Numerical_Results_Discussion} In this paper, the generic
power spectrum in (\ref{eq:Spectrum_h}) is used to simulate the
Rayleigh flat fading channel with non-isotropic scattering,
according to the spectral method\cite{IEEE_sw27:Acolatse03}. To
verify the accuracy of the derived formulas, we consider two types
of scattering environments: isotropic scattering and non-isotropic
scattering. For the non-isotropic scattering, we assume there are
three clusters of scatterers around the mobile station (MS), and
their parameters $[P_n, \kappa_n, \theta_n]$\footnote{These
parameters are defined in Appendix \ref{app:Channel_AoA_Models}, and
represent fractional contribution, azimuthal spread, and the
azimuthal location of the clusters in the two dimensional plane.}
are given by $[P_1, \kappa_1, \theta_1]=\left[0.45, 2,
\frac{\pi}{18}\right]$, $[P_2, \kappa_2, \theta_2]=\left[0.2, 20,
\frac{11\pi}{18}\right]$, and $[P_3, \kappa_3, \theta_3]=\left[0.35,
3, \frac{53\pi}{36}\right]$, respectively. In addition, in all the
simulation, the maximum Doppler frequency $f_m$ is set to 10Hz, and
$T_{\!s}=\frac{1}{20f_m}$ seconds.
\begin{figure}[tb]
\centering
\includegraphics[width=1.0\linewidth, clip=true, bb=-150 70 742 736]
{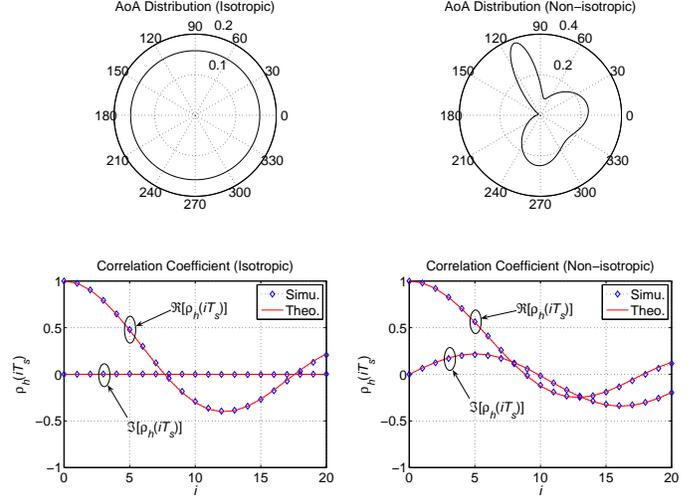} \caption{The AoA distributions
for two scattering environments and the corresponding channel
correlation coefficients.} \label{fig:AoA_Coeff_Plots_Iso_NonIso}
\end{figure}

The distributions of angle-of-arrival (AoA) for both scenarios and
the corresponding channel correlation coefficients
$\rho_h(iT_{\!s})$, $i\geq0$, are plotted in Fig.
\ref{fig:AoA_Coeff_Plots_Iso_NonIso}, where ``Simu.'' means
simulation, and ``Theo.'' refers to (\ref{eq:Coeff_h}), with
$f_m=10$ and $\tau=iT_{\!s}=\frac{i}{20f_m}$, $i=0, 1, 2, \cdots$.

In the following subsections, simulations are performed to verify
the ACF, correlation coefficient, LCR and AOD of IMI in both types
of environment at both low- and high-SNR regimes. For evaluating the
approximation accuracy of ACF and the correlation coefficient, we
set $\eta=-10$ dB for low SNR, and $\eta=30$ dB for high SNR. To
verify the LCR and AOD expressions, $\eta=0, 5, 10$ dB are the SNRs
considered.

\subsection{Isotropic Scattering}
\begin{figure}[tb]
\centering
\includegraphics[width=1.0\linewidth, clip=true, bb=-140 64 745 736]
{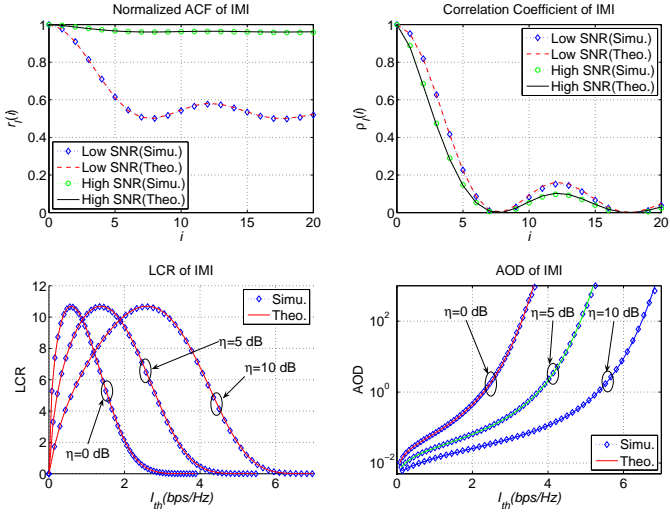} \caption{The ACF,
correlation coefficient, LCR, and AOD of IMI in a SISO system with
isotropic scattering.}
\label{fig:SISO_ACF_CorrCoeff_LCR_AOD_Plots_Iso}
\end{figure}

\begin{figure}[b]
\centering
\includegraphics[width=1.0\linewidth, clip=true, bb=-140 64 745 736]
{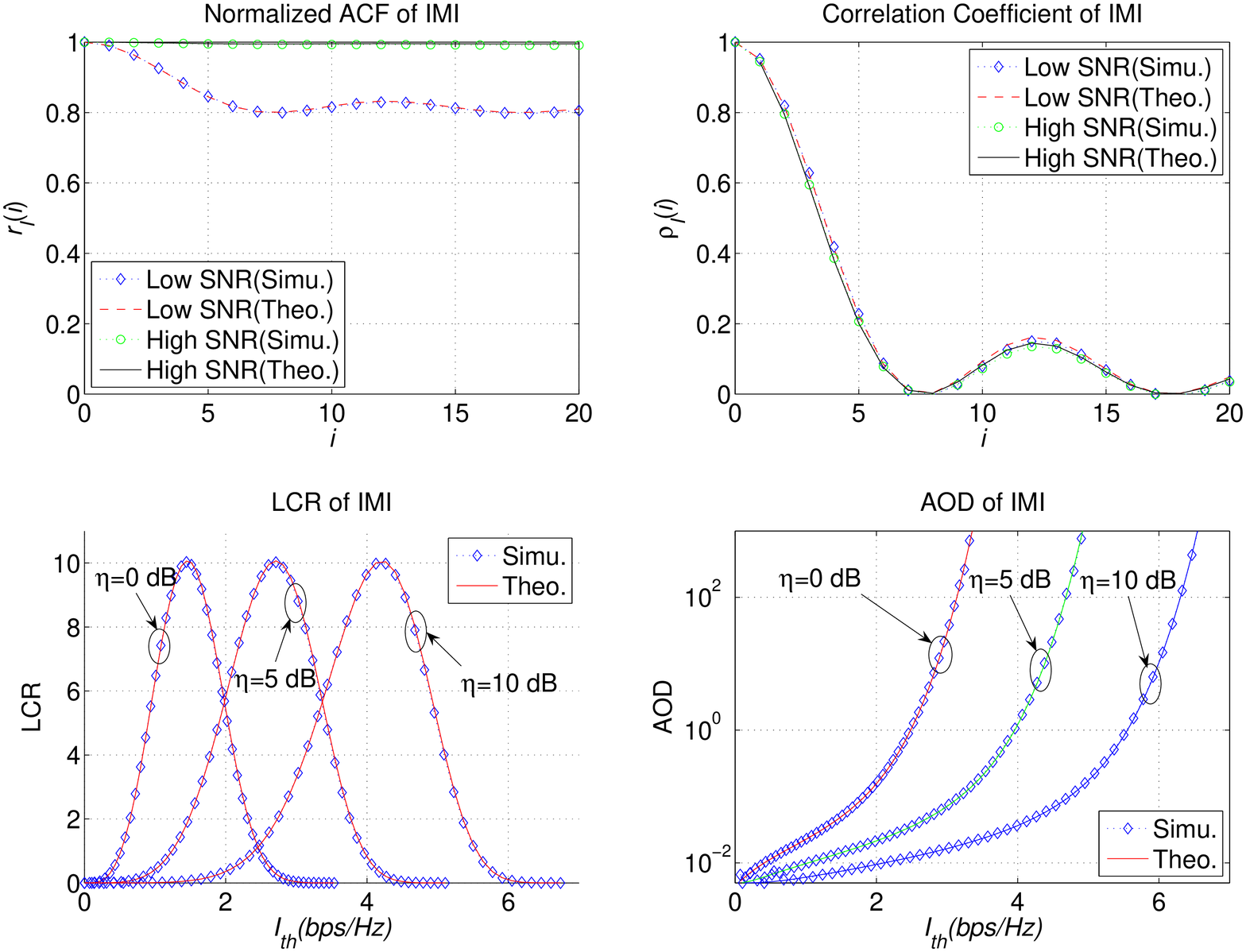} \caption{The ACF,
correlation coefficient, LCR, and AOD of IMI in a $2\times2$
OSTBC-MIMO system with isotropic scattering.}
\label{fig:OSTBC_MIMO_ACF_CorrCoeff_LCR_AOD_Plots_Iso}
\end{figure}

This is the Clarke's model, with uniform AoA. Simulations are
performed for both SISO and the $2\times2$ OSTBC-MIMO systems, with
the results shown in Figs.
\ref{fig:SISO_ACF_CorrCoeff_LCR_AOD_Plots_Iso} and
\ref{fig:OSTBC_MIMO_ACF_CorrCoeff_LCR_AOD_Plots_Iso}, respectively.

\subsection{Non-isotropic Scattering}
\begin{figure}[tb]
\centering
\includegraphics[width=1.0\linewidth, clip=true, bb=-140 64 745 736]
{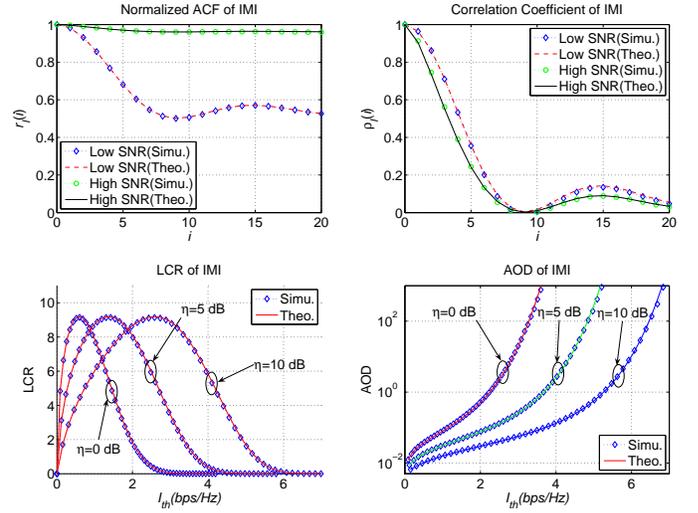} \caption{The ACF,
correlation coefficient, LCR, and AOD of IMI in a SISO system with
non-isotropic scattering.}
\label{fig:SISO_ACF_CorrCoeff_LCR_AOD_Plots_NonIso}
\end{figure}

\begin{figure}[b]
\centering
\includegraphics[width=1.0\linewidth, clip=true, bb=-140 64 745 736]
{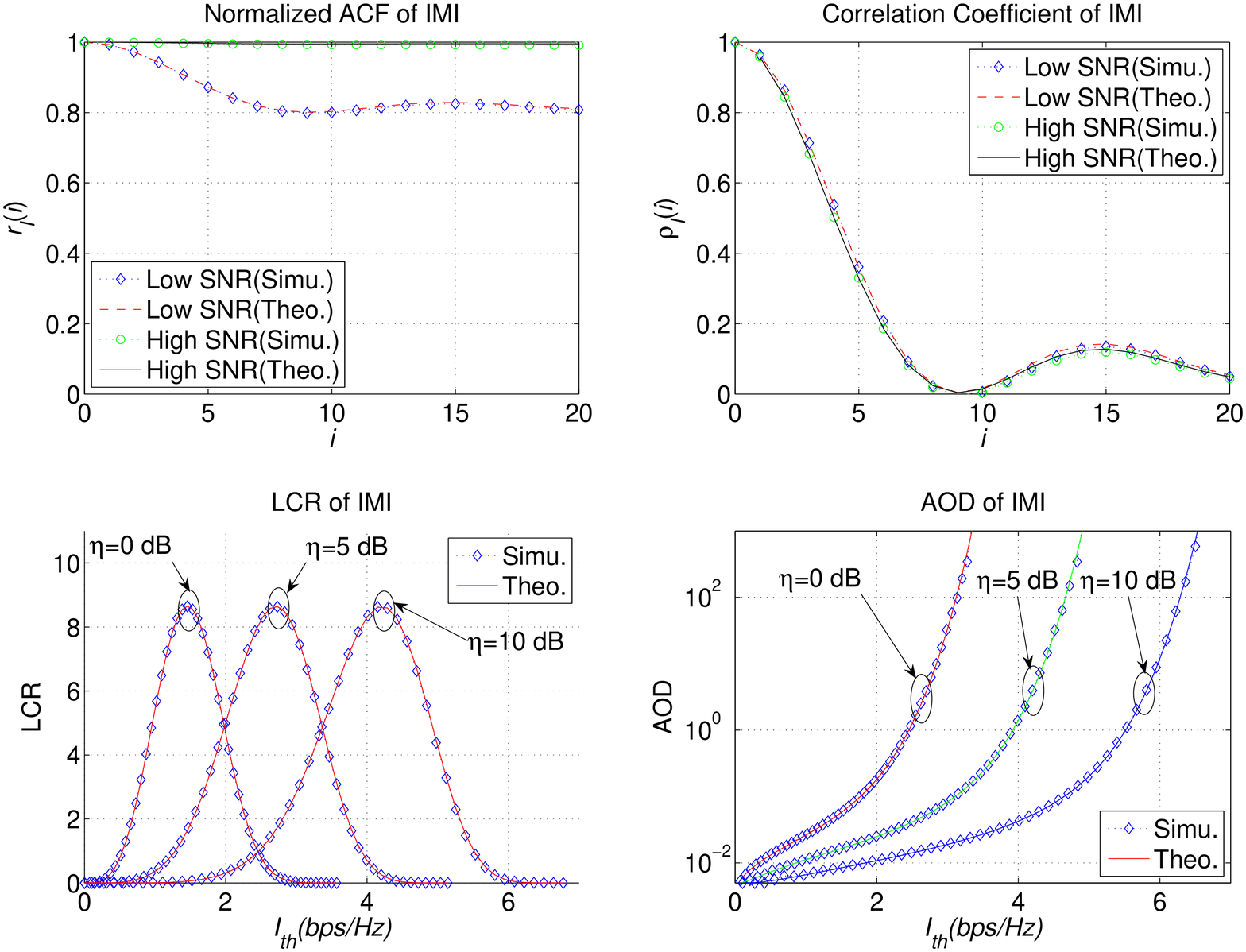} \caption{The
ACF, correlation coefficient, LCR, and AOD of IMI in a $2\times2$
OSTBC-MIMO system with non-isotropic scattering.}
\label{fig:OSTBC_MIMO_ACF_CorrCoeff_LCR_AOD_Plots_NonIso}
\end{figure}

This is a general case, with an arbitrary AoA distribution.
Simulations are also carried out for both SISO and the $2\times2$
OSTBC-MIMO systems, with the results presented in Figs.
\ref{fig:SISO_ACF_CorrCoeff_LCR_AOD_Plots_NonIso} and
\ref{fig:OSTBC_MIMO_ACF_CorrCoeff_LCR_AOD_Plots_NonIso},
respectively.

In Figs.
\ref{fig:SISO_ACF_CorrCoeff_LCR_AOD_Plots_Iso}-\ref{fig:OSTBC_MIMO_ACF_CorrCoeff_LCR_AOD_Plots_NonIso},
The upper left and right figures show the normalized approximate ACF
and the approximate correlation coefficient of the IMI
$\mathcal{I}_l$, respectively, in both low- and high-SNR regimes;
the lower left is the LCR of $\mathcal{I}_l$, and finally, the lower
right shows the AOD of $\mathcal{I}_l$. Moreover, in Figs.
\ref{fig:SISO_ACF_CorrCoeff_LCR_AOD_Plots_Iso} and
\ref{fig:SISO_ACF_CorrCoeff_LCR_AOD_Plots_NonIso}, for ACF,
theoretical values in the low- and high-SNR regimes are calculated
from (\ref{eq:NACF_LowSNR_SISO}) and (\ref{eq:NACF_HighSNR_SISO}),
respectively. For the correlation coefficient, theoretical values
come from (\ref{eq:Coeff_LowSNR_SISO}) and
(\ref{eq:Coeff_HighSNR_ISO_SISO}), for low- and high-SNR regimes,
respectively. The theoretical SISO LCR and AOD are given by
(\ref{eq:LCR_Up_DT_SISO}) and (\ref{eq:AOD_SISO}), respectively. In
Figs. \ref{fig:OSTBC_MIMO_ACF_CorrCoeff_LCR_AOD_Plots_Iso} and
\ref{fig:OSTBC_MIMO_ACF_CorrCoeff_LCR_AOD_Plots_NonIso}, for ACF,
theoretical values in the low- and high-SNR regimes are computed via
(\ref{eq:NACF_LowSNR_OSTBC}) and (\ref{eq:NACF_HighSNR_M2N2}),
respectively. For the correlation coefficient, theoretical values
come from (\ref{eq:Coeff_LowSNR_OSTBC}) and
(\ref{eq:Coeff_HighSNR_M2N2}) for low- and high-SNR regimes,
respectively. The theoretical MIMO LCR and AOD are given by
(\ref{eq:LCR_DT_Up}) and (\ref{eq:AOD_OSTBC}), respectively.

From Figs.
\ref{fig:SISO_ACF_CorrCoeff_LCR_AOD_Plots_Iso}-\ref{fig:OSTBC_MIMO_ACF_CorrCoeff_LCR_AOD_Plots_NonIso},
the following observations can be made.
\begin{itemize}
  \item By comparing Figs. \ref{fig:SISO_ACF_CorrCoeff_LCR_AOD_Plots_Iso}
and \ref{fig:SISO_ACF_CorrCoeff_LCR_AOD_Plots_NonIso} for the SISO
system, one can see larger peak values for LCR in the isotropic
scattering case, which means more fluctuations of IMI. On the other
hand, the AOD of IMI does not appear to be very sensitive to the
differences among the propagation examples considered. The same
conclusions apply to the $2\times2$ OSTBC-MIMO system, depicted in
Figs. \ref{fig:OSTBC_MIMO_ACF_CorrCoeff_LCR_AOD_Plots_Iso} and
\ref{fig:OSTBC_MIMO_ACF_CorrCoeff_LCR_AOD_Plots_NonIso}.
  \item Based on the results shown in Figs.
\ref{fig:SISO_ACF_CorrCoeff_LCR_AOD_Plots_Iso}-\ref{fig:OSTBC_MIMO_ACF_CorrCoeff_LCR_AOD_Plots_NonIso},
for LCR and AOD we can conclude that the theoretical expressions
perfectly match the simulations, in both propagation scenarios and
for both SISO and $2\times2$ OSTBC-MIMO systems. Moreover, the low-
and high-SNR approximations for ACF and correlation coefficient are
accurate in both types of systems and environments.
  \item Since in Figs.
\ref{fig:OSTBC_MIMO_ACF_CorrCoeff_LCR_AOD_Plots_Iso} and
\ref{fig:OSTBC_MIMO_ACF_CorrCoeff_LCR_AOD_Plots_NonIso}, the low-SNR
approximation for $\rho_\mathcal{I}(i)$ is very close to the
high-SNR approximation, one can conclude that
$\rho_\mathcal{I}(i)\approx\varrho_i^2$ for any SNR in the
OSTBC-MIMO systems with not-so-small number of antennas. This
statement is further supported by the results shown in Table
\ref{tab:Taylor_rho_I}, where for different values of $M\!N$, the
Taylor series of $\rho_\mathcal{I}(i)$ in
(\ref{eq:Coeff_HighSNR_OSTBC}) is given, as well as maximum values
$\max_{0\leq\varrho_i\leq1}|(\ref{eq:Coeff_LowSNR_OSTBC})-
(\ref{eq:Coeff_HighSNR_OSTBC})|$. From Table \ref{tab:Taylor_rho_I},
one can see that the larger the total number of antennas, the
smaller the difference between (\ref{eq:Coeff_LowSNR_OSTBC}) and
(\ref{eq:Coeff_HighSNR_OSTBC}). Moreover, even for $M\!N=2$, i.e., a
MIMO system with two Tx and one Rx antennas, or one Tx and two Rx
antennas, the difference is at most $0.075$, when using the
approximation $\rho_\mathcal{I}(i)\approx\varrho_i^2$.
   \item According to Figs.
\ref{fig:SISO_ACF_CorrCoeff_LCR_AOD_Plots_Iso} and
\ref{fig:SISO_ACF_CorrCoeff_LCR_AOD_Plots_NonIso}, there is small
gap between the SISO low- and high-SNR approximations of
$\rho_\mathcal{I}(i)$, which is $0.16$ at most, shown in Table
\ref{tab:Taylor_rho_I}. For the non-extreme SNR such as $10$ dB, the
non-approximate $\rho_\mathcal{I}(i)$ in
(\ref{eq:Coeff_InfSum_SISO}) can be used to calculate the
correlation coefficient. However, by evaluating
(\ref{eq:Coeff_InfSum_SISO}) and Monte Carlo simulations, we obtain
the following simple approximation
\begin{equation}\label{eq:Coeff_Approx_SISO}
\rho_\mathcal{I}(i)\approx\begin{cases}
\rho_i^2, & \eta\leq6.5\,\mathrm{dB},\\
\frac{\rho_i^2}{2}+\frac{3\Li_2\!\left(\varrho_i^2\right)}{\pi^2},&
6.5\,\mathrm{dB}<\eta\leq16\,\mathrm{dB},\\
\frac{6\Li_2\!\left(\varrho_i^2\right)}{\pi^2},&
\eta>16\,\mathrm{dB},
\end{cases}
\end{equation} and the approximation error is $0.055$ at most.
\end{itemize}

\section{Conclusion}
\label{sec:conclusion} In this paper, closed-form expressions for
the level crossing rate and average outage duration of the
instantaneous mutual information (IMI) in time-varying Rayleigh flat
fading channels are derived, as well as the autocorrelation function
and correlation coefficient of IMI, in SISO, MISO, SIMO and
OSTBC-MIMO systems. The analytical expressions, supported by Monte
Carlo simulations, provide useful qualitative and quantitative
information regarding the fluctuations of IMI. For example, as the
spread of the angle-of-arrival at the mobile receiver increases, the
crossing rate of IMI increases, according to our results, which
means faster fluctuation of IMI. Furthermore, IMI values become less
correlated. Quantification of the average outage duration of IMI is
another noteworthy outcome of this work. For example, with isotropic
scattering, the IMI of a SISO system, with a $10$ dB SNR and $10$ Hz
Doppler frequency, remains below $\mathcal{I}_\text{th}=6$ bps/Hz
for almost $7$ seconds, on average, i.e., an average outage duration
of $7$ seconds at the data rate $6$ bps/Hz.

\begin{table}[t]
\renewcommand{\arraystretch}{1.4}
\caption{Taylor Expansion of (\ref{eq:Coeff_HighSNR_OSTBC}) and the
Maximum Difference between (\ref{eq:Coeff_LowSNR_OSTBC}) and
(\ref{eq:Coeff_HighSNR_OSTBC}) for Different $M\!N$'s}
  \label{tab:Taylor_rho_I}
\centering
\begin{tabular}{|c||c|c|}
  \hline$M\!N$&$\text{Taylor Series of}\ (\ref{eq:Coeff_HighSNR_OSTBC})$&
  $\max_{0\leq\varrho_i\leq1}|(\ref{eq:Coeff_LowSNR_OSTBC})-
(\ref{eq:Coeff_HighSNR_OSTBC})|$ \\
  \hline
  \hline1&$0.608\varrho_i^2+0.152\varrho_i^4+\mathcal{O}(\varrho_i^6)$ &
  $0.16$ \\
  \hline2&$0.775\varrho_i^2+0.129\varrho_i^4+\mathcal{O}(\varrho_i^6)$ &
  $0.075$ \\
  \hline3&$0.844\varrho_i^2+0.106\varrho_i^4+\mathcal{O}(\varrho_i^6)$
  & $0.048$ \\
  \hline4&$0.881\varrho_i^2+0.088\varrho_i^4+\mathcal{O}(\varrho_i^6)$
  & $0.035$ \\
  \hline5&$0.904\varrho_i^2+0.075\varrho_i^4+\mathcal{O}(\varrho_i^6)$ &
  $0.027$ \\
  \hline$\vdots$ & $\vdots$ & $\vdots$ \\
  \hline16&$0.969\varrho_i^2+0.029\varrho_i^4+\mathcal{O}(\varrho_i^6)$
  & $0.008$ \\
  \hline$\vdots$ & $\vdots$ & $\vdots$ \\
  \hline64&$0.992\varrho_i^2+0.008\varrho_i^4+\mathcal{O}(\varrho_i^6)$
  & $0.002$ \\
  \hline
\end{tabular}
\end{table}
In this paper, we have not considered the spatially uncorrelated
general MIMO system with $M$ transmitters and $N$ receivers, where
all the subchannels are independent and identically distributed,
with the same temporal correlation function. The mutual information
in this case is given by\cite{IEEE_sw27:Telatar99_MIMO_Cap}
$\mathcal{I}_l=\log_2\det\!\left[\mathbf{I}_N+\frac{\eta}{M}\mathbf{H}(lT_{\!s})
\mathbf{H}^\dag(lT_{\!s})\right]$, where $\mathbf{H}(lT_{\!s})$ is
the channel matrix at time $t=lT_{\!s}$. However, for low SNRs, it
is straightforward to show that the MIMO IMI $\mathcal{I}_l$ is
quite similar to the OSTBC-MIMO IMI, which means that the
approximate normalized autocorrelation and correlation coefficient
in (\ref{eq:NACF_LowSNR_OSTBC}) and (\ref{eq:Coeff_LowSNR_OSTBC})
hold for more general MIMO systems as well\footnote{For the
correlation coefficient of MIMO IMI in the low-SNR regime, the same
result as $\rho_\mathcal{I}(i)\approx \varrho_i^2$ is derived in
\cite{IEEE_sw27:NanZhang05}.}. It is our ongoing work to derive
closed-form expressions for the IMI autocorrelation function,
correlation coefficient, LCR, and AOD of MIMO systems with any SNR,
and also some high-SNR approximations.

\appendices
\section{Channel and Angle-of-Arrival Models}
\label{app:Channel_AoA_Models} In a flat Rayleigh fading channel,
the lowpass complex envelope of the channel response $h(t)$ is a
zero-mean complex Gaussian random process, which can be represented
as
\begin{equation}\label{eq:Rayleigh_fading_channel}
h(t)=h_I(t)+\jmath h_Q(t)=\alpha(t)\exp[-\jmath\Phi(t)],
\end{equation} where the zero-mean real Gaussian random processes
$h_I(t)$ and $h_Q(t)$ are the real and imaginary parts of $h(t)$,
respectively. $\alpha(t)$ is the envelope of $h(t)$ and $\Phi(t)$ is
the phase of $h(t)$. At any time $t$, $\alpha(t)$ has a Rayleigh
distribution and $\Phi(t)$ is distributed uniformly over $[0,
2\pi)$. Without loss of generality, we assume the Rayleigh channel
has unit power, i.e., $\mathbb{E}[\alpha^2(t)]=1$.

To model the distribution of the angle-of-arrival (AoA) of waves
impinging either the base station (BS) or mobile station (MS), we
use multiple von Mises distributions.
\begin{figure}[b] \centering
\includegraphics[width=1.0\linewidth]{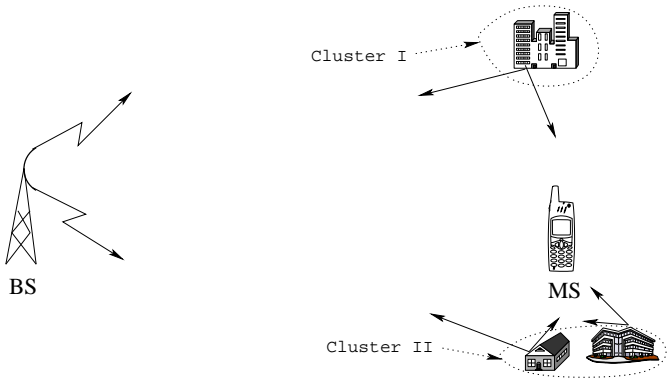}
\caption{The scattering environment around the MS with two clusters
of scatterers.} \label{fig:ScatteringEnv}
\end{figure}
The von Mises PDF has proven to be a flexible AoA model for both
BS\cite{IEEE_sw27:Abdi00_Corr_noniso} and
MS\cite{IEEE_sw27:Abdi02_AoA}. For example, in the macrocellular
environment shown in Fig.~\ref{fig:ScatteringEnv}, two von Mises
distributions are needed to model the two clusters of scatterers
around the MS. In general, we write the PDF of AoA in the azimuth
plane as the superposition of $K$ von Mises PDFs
\begin{equation}\label{eq:VonMises}
p(\theta)=\sum_{n=1}^KP_n\frac{\exp[\kappa_n\cos(\theta-\theta_n)]}{2\pi
I_0(\kappa_n)}, \quad \theta\in[0, 2\pi),
\end{equation} where $\theta_n$ is the mean AoA from the
$n^{\text{th}}$ cluster, $\kappa_n$ controls the width of the
$n^{\text{th}}$ cluster, and $P_n$ represents the contribution of
the $n^{\text{th}}$ cluster, such that $\sum_{n=1}^KP_n=1, 0<P_n
\leq 1$. If $\kappa_n=0, \forall n$, (\ref{eq:VonMises}) simplifies
to $p(\theta)=\frac{1}{2\pi}, \theta\in[0, 2\pi)$, which is the
Clarke's two-dimensional isotropic scattering model. The AoA PDF
given in (\ref{eq:VonMises}) is general enough to model AoA
distributions observed in practice. Furthermore, it provides
closed-form and mathematically-tractable expressions for the channel
correlation
functions~\cite{IEEE_sw27:Abdi00_Corr_noniso,IEEE_sw27:Abdi02_AoA}.

The channel correlation coefficient $\rho_h(\tau)$ and the power
spectrum of Rayleigh fading for the AoA model in (\ref{eq:VonMises})
are given by
\begin{align}\label{eq:Coeff_h}
\rho_h(\tau)&=\!E[h(t)h^*(t-\tau)],\\
&=\!\sum_{n=1}^K\!P_n\!\frac{I_0\!\left(\!\sqrt{\kappa_n^2
-4\pi^2f_m^2\tau^2+\jmath4\pi\kappa_n
f_m\tau\cos\theta_n}\right)}{I_0(\kappa_n)}\nonumber,
\end{align} and
\begin{multline}\label{eq:Spectrum_h}
    S_h(f)=\sum_{n=1}^KP_n\frac{\exp(\kappa_nf\cos\theta_n/f_m)}
    {\pi\sqrt{f_m^2-f^2}I_0(\kappa_n)}\\
    \times\cosh\!\left[\kappa_n\sqrt{1-(f/f_m)^2}\sin\theta_n\right]\!\!,
    |f|\leq f_m,
\end{multline} respectively, in which $f_m$ is the maximum Doppler
frequency, and $\cosh(z)=\frac{1}{2}(e^z+e^{-z})$ is the hyperbolic
cosine function. Equations (\ref{eq:Coeff_h}) and
(\ref{eq:Spectrum_h}) are derived by extending the results of
\cite{IEEE_sw27:Abdi02_AoA}, according to (\ref{eq:VonMises}). If
$\kappa_n=0, \forall n$, (\ref{eq:Coeff_h}) and
(\ref{eq:Spectrum_h}) simplify $\rho_h(\tau)=I_0(\jmath2\pi
f_m\tau)=J_0(2\pi f_m\tau)$ and
$S_h(f)=\frac{1}{2\pi\sqrt{f_m^2-f^2}}, |f|\leq f_m$, respectively,
which correspond to the Clarke's model.

Using (\ref{eq:Spectrum_h}) and the spectral method discussed in
\cite{IEEE_sw27:Acolatse03}, the discrete-time version of $h(t)$,
i.e., $h(lT_{\!s})$, is generated and used in
Sec.~\ref{sec:Numerical_Results_Discussion} to verify the accuracy
of the new closed-form expressions derived in this paper.

\section{Derivation of
(\ref{eq:ACF_InfSum_SISO}), (\ref{eq:NACF_InfSum_SISO}) and
(\ref{eq:Coeff_InfSum_SISO})} \label{app:SISO_IMI:sec:ACF_Coeff}
With (\ref{eq:jpdf_x1_x2_SISO}) and (\ref{eq:BesselI0_SeriesRep})
together, (\ref{eq:ACF_General_SISO}) can be written as
\begin{align}\label{eq:ACF_Expand_SISO}
\frac{r_\mathcal{I}(i)}{(\log_2e)^2}&=\int_0^\infty\!\!\!\!\int_0^\infty
\bigg[\!\ln(1+\eta x_1)\ln(1+\eta x_2)\nonumber\\
&\
\times\lambda_ie^{-\lambda_i(x_1+x_2)}\sum_{k=0}^{\infty}\frac{(\lambda_i\varrho_i)^{2k}
x_1^kx_2^k}{(k!)^2}\bigg]\mathrm{d}x_1\mathrm{d}x_2,\nonumber\\
&=\lambda_i\sum_{k=0}^{\infty}\bigg[\frac{(\lambda_i\varrho_i)^{2k}}
{(k!)^2}\!\!\int_0^\infty \!\!\!x_1^ke^{-\lambda_ix_1}\ln(1+\eta
x_1)\mathrm{d}x_1\nonumber\\
&\ \times\int_0^\infty \!\!\!x_2^ke^{-\lambda_ix_2}\ln(1+\eta
x_2)\mathrm{d}x_2\bigg],\nonumber\\
&=\lambda_i\!\sum_{k=0}^\infty\!\!\frac{(\lambda_i\varrho_i)^{2k}}
{(k!)^2}\!\!\left[\!\int_0^\infty\!\!\!x^ke^{-\lambda_ix}\ln(1\!+\!\eta
x)\mathrm{d}x\!\right]^2\!\!\!\!.
\end{align}

The exponential and logarithm functions can be represented by
Meijer's $G$ function as follows\cite[(11)]{IEEE_sw27:Adamchik90}
\begin{equation}\label{eq:MeijerG_for_Exp}
e^{-\lambda_ix}=G^{1, 0}_{0, 1}\!\left(\!\lambda_ix\left|\!\!
\begin{array}{c}
\cdot\\
0
\end{array}\right.
\!\!\right),
\end{equation} and
\begin{equation}\label{eq:MeijerG_for_ln}
\ln(1+\eta x)=G^{1, 2}_{2, 2}\!\left(\!\eta x\left|\!\!
\begin{array}{c}
1,1\\
1,0
\end{array}\right.
\!\!\right).
\end{equation}

Using the integral given in \cite[(21)]{IEEE_sw27:Adamchik90}, the
integral $\Xi(k,\eta,\lambda_i)=\int_0^\infty
x^ke^{-\lambda_ix}\ln(1+\eta x)\mathrm{d}x$ in
(\ref{eq:ACF_Expand_SISO}) can be expressed in the following form
\begin{equation}\label{eq:MeijerG_for_Integrand}
\begin{split}
\Xi(k,\eta,\lambda_i)&= \frac{1}{\eta^{k+1}}G_{2, 3}^{3,
1}\!\!\left(\!\frac{\lambda_i}{\eta}\left|\!\!\!
\begin{array}{c}
-(k\!+\!1),-k \\
-(k\!+\!1),-(k\!+\!1),0
\end{array}\right.
\!\!\!\!\right),\\
&=\frac{1}{\lambda_i^{k+1}}G_{2, 3}^{3,
1}\!\!\left(\!\frac{\lambda_i}{\eta}\left|\!\!\!
\begin{array}{c}
0,1 \\
0,0,k\!+\!1
\end{array}\right.
\!\!\!\!\right),
\end{split}
\end{equation} where the last ``='' comes from $z^kG_{p, q}^{m,
n}\!\!\left(z\left|\!\begin{array}{c}
(a_p) \\
(b_q)
\end{array}\right.
\!\!\!\right)=G_{p, q}^{m, n}\!\!\left(z\left|\!\begin{array}{c}
k+(a_p) \\
k+(b_q)
\end{array}\right.
\!\!\!\right)$\cite[pp.~521,
8.2.2.15]{IEEE_sw27:Prudnikov_Vol3_2nd}, in which $(a_p)=a_1, a_2,
\cdots, a_p$ and $(b_q)=b_1, b_2, \cdots, b_q$. By substituting
(\ref{eq:MeijerG_for_Integrand}) into (\ref{eq:ACF_Expand_SISO}), we
obtain (\ref{eq:ACF_InfSum_SISO}).

To obtain (\ref{eq:NACF_InfSum_SISO}) and
(\ref{eq:Coeff_InfSum_SISO}), we need to calculate
$\frac{\mathbb{E}\left[\mathcal{I}_l\right]}{\log_2e}=\mathbb{E}\left[\ln(1+\eta
x_1)\right]$ and
$\mathbb{E}\left[\mathcal{I}_l^2\right]=(\log_2\!e)^2\mathbb{E}\left[\ln^2(1+\eta
x_1)\right]$. With $p(x_1)=e^{-x_1}$, $x_1\geq0$, the characteristic
function of $\ln(1+\eta x_1)$ is given by
\begin{equation}\label{eq:CharacteristicFun_SISO}
\begin{split}
\xi_\mathcal{I}(t)&=\mathbb{E}\left[e^{\jmath t\ln(1+\eta
x_1)}\right]=\int_0^\infty
e^{-x_1}(1+\eta x_1)^{\jmath t}\,\mathrm{d}x_1,\\
&\hspace{-2em}\operatornamewithlimits{=}^{z=\frac{1}{\eta}+
x_1}\eta^{\jmath t}e^{\frac{1}{\eta}}\int_{\frac{1}{\eta}}^\infty
z^{\jmath t}e^{-z}\mathrm{d}z=\eta^{\jmath
t}e^{\frac{1}{\eta}}\Gamma\left(\jmath t+1, \frac{1}{\eta}\right).
\end{split}
\end{equation} The first and second
moments of $\ln(1+\eta x_1)$ are obtained as
\begin{equation}\label{eq:Mean_SISO}
\mathbb{E}\left[\ln(1+\eta
x_1)\right]=\frac{1}{\jmath}\left.\frac{\partial
\xi_\mathcal{I}(t)}{\partial t}
\right|_{t=0}=e^{\frac{1}{\eta}}\Gamma\left(0,\frac{1}{\eta}\right),
\end{equation} and
\begin{equation}\label{eq:2ndMoment_SISO}
\begin{split}
\mathbb{E}\left[\ln^2\!(1\!+\!\eta
x_1)\right]&=\frac{1}{\jmath^2}\!\!\left.\frac{\partial^2
\xi_\mathcal{I}(t)}{\partial t^2}
\right|_{t=0},\\
&=2e^{\frac{1}{\eta}}
G_{2,3}^{3,0}\!\!\left(\!\frac{1}{\eta}\!\left|\!
\begin{array}{c}
 1,1 \\
 0,0,0
\end{array}\right.
\!\!\!\right),
\end{split}
\end{equation} respectively. With
(\ref{eq:Mean_SISO}) and (\ref{eq:2ndMoment_SISO}), we easily obtain
(\ref{eq:NACF_InfSum_SISO}) and (\ref{eq:Coeff_InfSum_SISO}).
\section{Derivation of (\ref{eq:NACF_HighSNR_SISO}) and
(\ref{eq:Coeff_HighSNR_SISO})}
\label{app:SISO_IMI:sec:ACF_Coeff_High_SNR} By plugging
(\ref{eq:Xi_H}) into (\ref{eq:ACF_XiH_HighSNR_SISO}) we obtain
\begin{equation}\label{eq:ACF_S012_HighSNR_SISO}
\begin{split}
\frac{r_\mathcal{I}(i)}{(\log_2e)^2}&\approx\frac{1}{\lambda_i}\!
\sum_{k=0}^{\infty}\varrho_i^{2k}\left(\ln\frac{\eta}{\lambda_i
\gamma}+{H}_{\!k}\right)^2,\\
&\hspace{-4em}=\frac{1}{\lambda_i}\left[S_2(\varrho_i^2)+2S_1(\varrho_i^2)
\ln\frac{\eta}{\lambda_i\gamma}+S_0(\varrho_i^2)
\ln^2\frac{\eta}{\lambda_i\gamma}\right],
\end{split}
\end{equation} where\cite{IEEE_sw27:Neumann91_IMA}
\begin{subequations}\label{eq:S012}
\begin{align}
S_0(\varrho_i^2)&=\sum_{k=0}^\infty\varrho_i^{2k},\nonumber\\
&=\lambda_i,\label{eq:S0}\\
S_1(\varrho_i^2)&=\sum_{k=0}^\infty{H}_{\!k}
\varrho_i^{2k},\nonumber\\
&=\lambda_i\ln\lambda_i,\label{eq:S1}\\
S_2(\varrho_i^2)&=\sum_{k=0}^\infty{H}_{\!k}^2
\varrho_i^{2k},\nonumber\\
&=\lambda_i\left[\Li_2\!\left(\varrho_i^2\right)
+\ln^2\lambda_i\right]\label{eq:S2}.
\end{align}
\end{subequations}

Substitution of (\ref{eq:S012}) into
(\ref{eq:ACF_S012_HighSNR_SISO}) yields
\begin{equation}\label{eq:ACF_HighSNR_SISO}
\frac{r_\mathcal{I}(i)}{(\log_2e)^2}=\Li_2\!\left(\varrho_i^2\right)
+\ln^2\frac{\eta}{\gamma}.
\end{equation} After normalization, i.e.,
$\tilde{r}_\mathcal{I}(i)
=\frac{r_\mathcal{I}(i)}{r_\mathcal{I}(0)}$, where
$r_\mathcal{I}(0)\approx(\log_2e)^2\left(
\frac{\pi^2}{6}+\ln^2\frac{\eta}{\gamma}\right)$,
(\ref{eq:ACF_HighSNR_SISO}) reduces to (\ref{eq:NACF_HighSNR_SISO}).

The correlation coefficient is calculated as follows
\begin{equation}\label{app:eq:Coeff_HighSNR_SISO}
\begin{split}
\rho_\mathcal{I}(i)&\approx\frac{r_\mathcal{I}(i)-
\left\{\left(\log_2e\right)\mathbb{E}[\ln(\eta
x_1)]\right\}^2}{r_\mathcal{I}(0)-\left\{\left(\log_2e\right)\mathbb{E}[\ln(\eta
x_1)]\right\}^2},\\
&\approx\frac{\Li_2\!\left(\varrho_i^2\right)+\ln^2\frac{\eta}{\gamma}-
\ln^2\frac{\eta}{\gamma}}{\frac{\pi^2}{6}+\ln^2\frac{\eta}{\gamma}-
\ln^2\frac{\eta}{\gamma}},\\
&=\frac{6\Li_2\!\left(\varrho_i^2\right)}{\pi^2},
\end{split}
\end{equation} where the second line is from the fact that
$\mathbb{E}\left[\ln(\eta x_1)\right]=\ln\eta+\int_0^\infty
e^{-x_1}\ln x_1\,\mathrm{d}x_1=\ln\frac{\eta}{\gamma}$ using
$\int_0^\infty e^{-x_1}\ln
x_1\,\mathrm{d}x_1=-\ln\gamma$\cite[pp.~602,
4.331.1]{IEEE_sw27:RyzhikBook_5th}.
\section{Derivation of Some Results in Sec. \ref{sec:OSTBC}}
\label{app:Derivation_for_OSTBC}
\subsection{Derivation of (\ref{eq:NACF_OSTBC}) and
(\ref{eq:Coeff_OSTBC})} \label{app:Deriv_ACF_Coeff_OSTBC}
Substitution of (\ref{eq:Def_IMI_OSTBC}) and
(\ref{eq:jpdf_y1_y2_OSTBC}) into (\ref{eq:Def_ACF_SISO}) results in
\begin{align}\label{eq:ACF_Expand_OSTBC}
\frac{r_\mathcal{I}(i)}{(\log_2e)^2}&=\int_0^\infty\int_0^\infty
\Bigg[\ln\left(1+\frac{\eta}{M} y_1\right)\ln\left(1+\frac{\eta}{M}
y_2\right)\nonumber\\
&\hspace{1em}\times\frac{\lambda_i^{M\!N}
(y_1y_2)^{M\!N-1}e^{-\lambda_i(y_1+y_2)}}
{(M\!N-1)!}\nonumber\\
&\hspace{1em}\times\sum_{k=0}^{\infty}\frac{(\lambda_i\varrho_i)^{2k}
(y_1y_2)^k}{k!(M\!N+k-1)!}\Bigg]\mathrm{d}y_1\mathrm{d}y_2,\nonumber\\
&=\frac{\lambda_i^{M\!N}}{(M\!N-1)!}
\sum_{k=0}^{\infty}\Bigg[\frac{(\lambda_i\varrho_i)^{2k}}
{k!(M\!N+k-1)!}\nonumber\\
&\hspace{1em}\times\int_0^\infty
y_1^{k+M\!N-1}e^{-\lambda_iy_1}\ln\left(1+\frac{\eta}{M}
y_1\right)\mathrm{d}y_1
\nonumber\\
&\hspace{1em}\times\int_0^\infty
y_2^{k+M\!N-1}e^{-\lambda_iy_2}\ln\left(1+\frac{\eta}{M}
y_2\right)\mathrm{d}y_2\Bigg],\nonumber\\
&\hspace{-1em}=\frac{\lambda_i^{M\!N}}{(M\!N-1)!}
\sum_{k=0}^\infty\Bigg\{\frac{(\lambda_i\varrho_i)^{2k}}
{k!(M\!N+k-1)!}\nonumber\\
&\hspace{-1em}\times\left[\int_0^\infty
y_1^{k+M\!N-1}e^{-\lambda_iy_1}\ln\left(1+\frac{\eta}{M}
y_1\right)\mathrm{d}y_1\right]^2\Bigg\},\nonumber\\
&\hspace{-4em}=\frac{\lambda_i^{-M\!N}}{(M\!N-1)!}
\sum_{k=0}^\infty\frac{\varrho_i^{2k} \left[G_{2, 3}^{3,
1}\!\!\left(\!\frac{M\lambda_i}{\eta}\!\left|\!\!
\begin{array}{c}
0,1 \\
0,0,k\!+\!M\!N
\end{array}\right.
\!\!\!\!\right)\right]^2} {k!(M\!N+k-1)!}.
\end{align}

For calculating the normalized ACF and the correlation coefficient,
we need the first and second moments of $\mathcal{I}_l$, i.e.,
$\mathbb{E}\left[\ln\left(1+\frac{\eta}{M} y_1\right)\right]$ and
$\mathbb{E}\left[\ln^2\left(1+\frac{\eta}{M} y_1\right)\right]$.
Using the integral results given in
(\ref{eq:MeijerG_for_Integrand}), we have the first moment as
\begin{align}\label{eq:Mean_OSTBC}
\mathbb{E}\left[\ln\left(1+\frac{\eta}{M}
y_1\right)\right]&=\int_0^\infty\frac{y_1^{M\!N-1}e^{-y_1}}
{(M\!N-1)!}\ln\left(1+\frac{\eta}{M}y_1\right)\mathrm{d}y_1,\nonumber\\
&=\frac{G_{2,3}^{3,1}\!\left(\!\frac{M}{\eta}\!\left|
\begin{array}{c}
 0,1 \\
 0,0,M\!N
\end{array}
\right.\!\!\right)}{(M\!N\!-\!1)!}.
\end{align}

The second moment is given in (\ref{eq:2ndMoment_OSTBC}), where the
third ``='' is based on the integral identity 4.358.1 in
\cite[pp.~607]{IEEE_sw27:RyzhikBook_5th}.
\begin{figure*}[!t]
\newcounter{eq_2ndMoment_OSTBC}
\normalsize \setcounter{eq_2ndMoment_OSTBC}{\value{equation}}
\setcounter{equation}{68}
\begin{equation}\label{eq:2ndMoment_OSTBC}
\begin{split}
\mathbb{E}\left[\ln^2\left(1+\frac{\eta}{M}
y_1\right)\right]&=\int_0^\infty\frac{y_1^{M\!N-1}e^{-y_1}}
{(M\!N-1)!}\ln^2\left(1+\frac{\eta}{M}y_1\right)\mathrm{d}y_1,\\
&\operatornamewithlimits{=}^{z=1+\frac{\eta}{M}
y_1}\frac{e^\frac{M}{\eta}\sum_{j=0}^{M\!N-1}{M\!N-1 \choose
j}(-1)^{M\!N-1-j} \int_1^\infty
z^je^{-\frac{M}{\eta}z}\ln^2z\mathrm{d}z}{(M\!N\!-\!1)!\left(\frac{\eta}{M}\right)^{M\!N}},\\
&=\frac{e^\frac{M}{\eta}\sum_{j=0}^{M\!N-1}{M\!N-1 \choose
j}(-1)^{M\!N-1-j}\frac{\partial^2} {\partial
\nu^2}\!\!\left.\left[\!\left(\frac{M}{\eta}\right)^{-\nu}
\Gamma\!\left(\nu,\frac{M}{\eta}\right)\!\right]\!
\right|_{\nu=j+1}}{(M\!N\!-\!1)!\left(\frac{\eta}{M}\right)^{M\!N}},\\
&=\frac{2e^\frac{M}{\eta}{\displaystyle\sum_{j=0}^{M\!N-1}}{M\!N-1
\choose
j}(-1)^{M\!N-1-j}G_{3,4}^{4,0}\!\left(\!\frac{M}{\eta}\!\left|
\begin{array}{c}
 -\!j,-\!j,-\!j \\
 0,-\!j\!-\!1,-\!j\!-\!1,-\!j\!-\!1
\end{array}
\right.\!\!\!\!\right)}{(M\!N\!-\!1)!\left(\frac{\eta}{M}\right)^{M\!N}}.
\end{split}
\end{equation}
\setcounter{equation}{\value{eq_2ndMoment_OSTBC}} \hrulefill
\vspace*{4pt}
\end{figure*}
\subsection{Derivation of (\ref{eq:NACF_LowSNR_OSTBC}) and
(\ref{eq:Coeff_LowSNR_OSTBC})}
\label{app:Deriv_ACF_Coeff_OSTBC_MIMO_Low_SNR} For calculating the
low-SNR approximation for the normalized ACF and the correlation
coefficient, the following lemma is necessary.
\begin{lem}\label{lem:Sum_of_InfSeries}
If $|t|<1$ and $p$ is a non-negative integer, then the following
identity holds.
\begin{equation}
\sum_{k=0}^\infty\frac{(k+p)!(k+p)} {k!}t^k=\frac{(p+t)p!}
{(1-t)^{p+2}}.
\end{equation}
\addtocounter{equation}{1}
\end{lem}
\begin{proof}
For $p=0$, obviously it holds\cite[pp.~8,
0.231.2]{IEEE_sw27:RyzhikBook_5th}.

For $p\geq1$ we have
\begin{equation*}
\begin{split}
\sum_{k=0}^\infty\frac{(k+p)!(k+p)} {k!}t^k&=\frac{\partial^{p-1}}
{\partial
t^{p-1}}\left[t^{p-1}\sum_{k=0}^\infty (k+p)^2t^k\right],\\
&\hspace{-8em}=\frac{\partial^{p-1}} {\partial
t^{p-1}}\left[\frac{\partial} {\partial
t}\sum_{k=0}^\infty(k+p)t^{k+p} \right]=\frac{\partial^p} {\partial
t^p}\left[t\frac{\partial}
{\partial t}\sum_{k=0}^\infty t^{k+p} \right],\\
&\hspace{-8em}=\frac{\partial^p} {\partial
t^p}\left[t\frac{\partial} {\partial t}\frac{t^p}{1-t}
\right]=\frac{\partial^p} {\partial
t^p}\left[\frac{t^p(p-pt+t)}{(1-t)^2}\right]=\frac{(p+t)p!}
{(1-t)^{p+2}},
\end{split}
\end{equation*} where the fourth ``='' is from $\sum_{k=0}^\infty t^{k+p}
=\frac{t^p}{1-t}$, when $|t|<1$ and $p$ is an positive integer.
\end{proof}

Using the low-SNR approximation, similar to
(\ref{eq:Xi_LowHighSNR_SISO}) in the SISO case, with $x$, $k$, and
$\eta$ replaced by $y_1$, $k+M\!N-1$, and $\eta/M$, respectively,
(\ref{eq:ACF_Expand_OSTBC}) reduces to
\begin{align}\label{eq:ACF_LowSNR_OSTBC}
\frac{r_\mathcal{I}(i)}{(\log_2e)^2}
&\approx\frac{\eta^2\lambda_i^{M\!N}}{M^2(M\!N-1)!}
\sum_{k=0}^\infty\Bigg\{\frac{(\lambda_i\varrho_i)^{2k}}
{k!(M\!N+k-1)!}\nonumber\\
&\hspace{1em}\times\left[\int_0^\infty
y_1^{k+M\!N}e^{-\lambda_iy_1}\mathrm{d}y_1\right]^2\Bigg\},\nonumber\\
&\hspace{-3em}=\frac{\eta^2\lambda_i^{M\!N}}{M^2(M\!N-1)!}
\sum_{k=0}^\infty\frac{(\lambda_i\varrho_i)^{2k}}
{k!(M\!N+k-1)!}\left[\frac{(k+M\!N)!}{\lambda_i^{k+M\!N+1}}\right]^2,\nonumber\\
&\hspace{-3em}=\frac{\eta^2}{M^2(M\!N-1)!\lambda_i^{M\!N+2}}
\sum_{k=0}^\infty\frac{(k+M\!N)!(k+M\!N)}
{k!}(\varrho_i^2)^k,\nonumber\\
&\hspace{-3em}=\frac{\eta^2 N(M\!N+\varrho_i^2)}{M},
\end{align} where the last ``='' is from Lemma
\ref{lem:Sum_of_InfSeries}. Therefore we have the approximate
normalized ACF as given in (\ref{eq:NACF_LowSNR_OSTBC}). For the
approximate correlation coefficient we obtain
\begin{equation}\label{app:eq:Coeff_LowSNR_OSTBC}
\begin{split}
\rho_\mathcal{I}(i)&\approx\frac{r_\mathcal{I}(i)-
\left\{\left(\log_2e\right)\mathbb{E}\left[\frac{\eta}{M}y\right]
\right\}^2}{r_\mathcal{I}(0)-\left\{\left(\log_2e\right)\mathbb{E}
\left[\frac{\eta}{M}y\right]\right\}^2},\\
&=\frac{\frac{\eta^2 N(M\!N+\varrho_i^2)}{M}-\eta^2N^2}
{\frac{\eta^2 N(M\!N+1)}{M}-\eta^2N^2}=\varrho_i^2,
\end{split}
\end{equation}
where $\mathbb{E}[y_1]=M\!N$\cite[pp. 14,
(2.35)]{IEEE_sw27:SimonBook02} is used in the second line of
(\ref{app:eq:Coeff_LowSNR_OSTBC}).
\subsection{Derivation of (\ref{eq:NACF_HighSNR_OSTBC}) and
(\ref{eq:Coeff_HighSNR_OSTBC})}
\label{app:Deriv_ACF_Coeff_OSTBC_MIMO_High_SNR} For the high-SNR
regime, the approximation of $\frac{r_\mathcal{I}(i)}{(\log_2e)^2}$
is detailed in (\ref{eq:ACF_HighSNR_OSTBC}), where a high-SNR
approximation similar to (\ref{eq:Xi_LowHighSNR_SISO}) in the SISO
case is used, and $R_j(\varrho_i^2)$, $j=0, 1, 2$, are given in
(\ref{eq:R0}), (\ref{eq:R1}), and (\ref{eq:R2}), respectively.
\begin{figure*}[!t]
\newcounter{eq:ACF_HighSNR_OSTBC}
\normalsize \setcounter{eq:ACF_HighSNR_OSTBC}{\value{equation}}
\setcounter{equation}{76}
\begin{align}\label{eq:ACF_HighSNR_OSTBC}
\frac{r_\mathcal{I}(i)}{(\log_2e)^2}&=\frac{\lambda_i^{M\!N}}{(M\!N-1)!}
\sum_{k=0}^\infty\frac{(\lambda_i\varrho_i)^{2k}}
{k!(M\!N+k-1)!}\left[\int_0^\infty
y_1^{k+M\!N-1}e^{-\lambda_iy_1}\ln\left(1+\frac{\eta}{M}
y_1\right)\mathrm{d}y_1\right]^2,\nonumber\\
&\approx\frac{\lambda_i^{M\!N}}{(M\!N-1)!}
\sum_{k=0}^\infty\frac{(\lambda_i\varrho_i)^{2k}}
{k!(M\!N+k-1)!}\left[\int_0^\infty
y_1^{k+M\!N-1}e^{-\lambda_iy_1}\ln\left(\frac{\eta}{M}
y_1\right)\mathrm{d}y_1\right]^2,\nonumber\\
&=\frac{\lambda_i^{M\!N}}{(M\!N-1)!}
\sum_{k=0}^\infty\frac{(\lambda_i\varrho_i)^{2k}}
{k!(M\!N+k-1)!}\left[\frac{(k+M\!N-1)!}{\lambda_i^{k+M\!N}}
\left(\ln\frac{\eta}{M\lambda_i\gamma}
+{H}_{\!k+M\!N-1}\right)\right]^2,\nonumber\\
&=\frac{1}{(M\!N-1)!\lambda_i^{M\!N}}
\sum_{k=0}^\infty\frac{(k+M\!N-1)!(\varrho_i^2)^k}{k!}
\left(\ln\frac{\eta}{M\lambda_i\gamma}
+{H}_{\!k+M\!N-1}\right)^2,\nonumber\\
&=\frac{1}{(M\!N-1)!\lambda_i^{M\!N}}\left[R_2(\varrho_i^2)
+2R_1(\varrho_i^2) \ln\frac{\eta}{M\lambda_i\gamma}+R_0(\varrho_i^2)
\ln^2\frac{\eta}{M\lambda_i\gamma}\right].
\end{align}
\setcounter{equation}{\value{eq:ACF_HighSNR_OSTBC}} \hrulefill
\vspace*{4pt}
\end{figure*}
In what follows, we focus on how to calculate the three infinite
summations $R_j(t), j=0,1,2, |t|<1$, introduced in
(\ref{eq:ACF_HighSNR_OSTBC}). First, we consider $R_0(t)$
\begin{align}
R_0(t)&=\sum_{k=0}^\infty\! \frac{(k\!+\!M\!N\!-\!1)!t^k}{k!}
=\sum_{k=0}^\infty\frac{\partial^{M\!N-1}}{\partial
t^{M\!N-1}} t^{k+M\!N-1},\nonumber\\
&\hspace{-3em}=\frac{\partial^{M\!N-1}}{\partial
t^{M\!N-1}}\sum_{k=0}^\infty
t^{k+M\!N-1}=\frac{\partial^{M\!N-1}}{\partial
t^{M\!N-1}}\left(\sum_{k=0}^\infty t^k-\sum_{k=0}^{M\!N-2}
t^k\right),\nonumber\\
&\hspace{-3em}=\frac{\partial^{M\!N-1}}{\partial
t^{M\!N-1}}\sum_{k=0}^\infty
t^k\operatornamewithlimits{=}^{|t|<1}\frac{\partial^{M\!N-1}S_0(t)}{\partial
t^{M\!N-1}}=\frac{(M\!N-1)!}{(1-t)^{M\!N}},
\end{align} where we use $S_0(t)=\frac{1}{1-t}$ in (\ref{eq:S0}).
Therefore we have
\begin{equation}\label{eq:R0}
R_0(\varrho_i^2)=\sum_{k=0}^\infty
\frac{(k+M\!N-1)!}{k!}\varrho_i^{2k}=
\frac{(M\!N-1)!}{(1-\varrho_i^2)^{M\!N}}.
\end{equation}
With the same reasoning we obtain
\begin{equation}
R_j(t)=\frac{\partial^{M\!N-1}S_j(t)}{\partial t^{M\!N-1}}, j=1,2,
\end{equation} where $S_j(t),
j=1,2$, are given by (\ref{eq:S1}) and (\ref{eq:S2}), respectively.
Hence
\begin{align}
R_1(\varrho_i^2)&=\sum_{k=0}^\infty\frac{(k+M\!N-1)!}{k!}{H}_{\!k+M\!N-1}
\varrho_i^{2k},\nonumber\\
&=\frac{\partial^{M\!N-1}} {\partial
t^{M\!N-1}}\left.\left[-\frac{\ln(1-t)}{(1-t)}\right]
\right|_{t=\varrho_i^2},\nonumber\\
&=\frac{(M\!N-1)!\left[{H}_{\!M\!N-1}-\ln(1-\varrho_i^2)\right]}
{(1-\varrho_i^2)^{M\!N}},\label{eq:R1}\\
R_2(\varrho_i^2)&=\sum_{k=0}^\infty\frac{(k+M\!N-1)!}{k!}{H}_{\!k+M\!N-1}^2
\varrho_i^{2k},\nonumber\\
&\hspace{-1em}=\frac{\partial^{M\!N-1}} {\partial
t^{M\!N-1}}\left.\left[\frac{\Li_2(t)+\ln^2(1-t)}{(1-t)}\right]
\right|_{t=\varrho_i^2}\label{eq:R2}.
\end{align}
\addtocounter{equation}{1}

The second moment of $\ln\left(\frac{\eta}{M}y\right)$ is given by
\begin{align}\label{eq:2ndMoment_HighSNR_OSTBC}
\frac{r_\mathcal{I}(0)}{(\log_2e)^2}&\approx\int_0^\infty
\frac{y_1^{M\!N-1}e^{-y_1}}{(M\!N-1)!}\ln^2\frac{\eta}{M}y_1\mathrm{d}y_1,
\nonumber\\
&\operatornamewithlimits{=}^{z=\frac{\eta}{M}y_1}
\frac{\left(\frac{M}{\eta}\right)^{M\!N}}{(M\!N-1)!}\int_0^\infty
z^{M\!N-1}e^{-\frac{M}{\eta}z}\ln^2z\mathrm{d}z,\nonumber\\
&=\left({H}_{\!M\!N-1}+\ln\frac{\eta}{M\gamma}\right)^2+\zeta(2,M\!N),
\end{align} in which the last ``='' is from 4.358.2
\cite[pp.~607]{IEEE_sw27:RyzhikBook_5th}, where in
\cite{IEEE_sw27:RyzhikBook_5th} $\psi(q)={H}_{\!q-1}-\ln\gamma$,
with $q$ as a positive integer\cite[pp.~952,
8.365.4]{IEEE_sw27:RyzhikBook_5th}. Therefore, we obtain
(\ref{eq:NACF_HighSNR_OSTBC}) for the normalized ACF. With
\begin{equation}
\begin{split}
\mathbb{E}\left[\ln\left(\frac{\eta}{M}y\right)\right]&=
\int_0^\infty\frac{y_1^{M\!N-1}e^{-y_1}}{(M\!N-1)!}
\ln\left(\frac{\eta}{M}y_1\right)\mathrm{d}y_1,\\
&=\ln\frac{\eta}{M\gamma}+{H}_{\!M\!N-1},
\end{split}
\end{equation} where the last ``='' is from
(\ref{eq:Xi_H}), we have (\ref{eq:Coeff_HighSNR_OSTBC}) for the
correlation coefficient.
\bibliographystyle{IEEEtran}
\bibliography{IEEEabrv,IEEE_sw27}
\end{document}